\documentstyle[aps,multicol,epsfig,prbbib,floats]{revtex}

\begin{document}
\draft
\twocolumn[\hsize\textwidth\columnwidth\hsize\csname
@twocolumnfalse\endcsname
\title{Magnetotransport in the Normal State of 
La$_{1.85}$Sr$_{0.15}$Cu$_{1-y}$Zn$_y$O$_4$ Films}
\author{A. Malinowski and Marta Z. Cieplak }
\address{Department of Physics and Astronomy, Rutgers University, Piscataway,
NJ 08855, USA\\
and Institute of Physics, Polish Academy of Sciences, 02 668 Warsaw, Poland}
\author{S. Guha, Q. Wu, B. Kim, A. Krickser, and A. Perali}
\address{Department of Physics and Astronomy, Rutgers University, Piscataway,
NJ 08855, USA}
\author{K. Karpi\'{n}ska and M. Berkowski}
\address{Institute of Physics, Polish Academy of Sciences, 02 668 Warsaw, Poland}
\author{C. H. Shang}
\address{Department of Physics and Astronomy, The Johns Hopkins University, 
Baltimore, Md 21218, USA}
\author{P. Lindenfeld}
\address{Department of Physics and Astronomy, Rutgers University, Piscataway, 
NJ 08855, USA}

\maketitle

\begin{abstract}
We have studied the magnetotransport properties in the normal state for a series 
of La$_{1.85}$Sr$_{0.15}$Cu$_{1-y}$Zn$_y$O$_4$ films with values of
$y$, between 0 and 0.12. A variable degree of 
compressive or tensile strain results from the lattice mismatch between 
the substrate and the film, and affects the transport properties
differently from the influence of the zinc impurities.
In particular, the orbital magnetoresistance (OMR) varies with $y$ but
is strain--independent. The relations for the resistivity 
($\rho = \rho_0 + AT$) and the Hall angle ($\cot{\Theta_H}=\alpha
T^{2}+C$), and the proportionality between the OMR and $\tan^2{\Theta_H}$
($\Delta\rho /\rho = \zeta \tan^2 {\Theta_H}$) are followed above about
70 K. We have been able to separate the strain and impurity effects by
rewriting the last two of these relations as $\cot{\Theta_H}/\alpha =
T^2 + C/\alpha$ and $\Delta \rho/\rho = (\zeta/\alpha^2)(\alpha^2
tan^2~{\Theta_H})$, where each term is strain--independent and depends on
$y$ only. We also find that changes in the lattice constants give rise
to closely the same fractional changes in $A$, $C$, and $\alpha$, while
$\rho_0$ is, in addition, increased by changes in the microstructure.
The OMR is more strongly suppressed by the addition of impurities than
$\tan^2{\Theta_H}$, so that $\zeta$ decreases as $y$ increases. We
conclude that the relaxation rate that governs the Hall effect is not
the same as for the magnetoresistance. We also suggest a correspondence
between the transport properties and the opening of the pseudogap at a
temperature which changes when the La--Sr ratio changes, but does not
change with the addition of the zinc impurities. Several theoretical
models seem to be in conflict with our results. Some recent
ones may be more compatible, but have not been carried sufficiently far
for a detailed comparison. 
\end{abstract}

]

\section {Introduction}

One of the most puzzling aspects of the high-$T_c$ superconductors is their
behavior in the normal state which appears to be distinctly different from
that of any other metal.\cite{iye} All normal--state properties of
high-$T_c$ superconductors  
display anomalous temperature dependences. In particular, their electrical charge 
transport is difficult to understand on the basis of a simple Drude model,
in which the temperature dependence of various scattering 
processes is described with a single relaxation time $\tau$. 
In high--$T_c$ materials the in--plane resistivity, $\rho_{xx} \equiv
\rho$, follows the relation \cite{gf,takagi}

\begin{equation}
\rho =\rho_{0}+AT.
\label{resis}
\end{equation}

The Hall coefficient, 
$R_H = {{\sigma}_{xy}}/(H{{\sigma}_{xx}}{{\sigma}_{yy}})$ 
(where ${\sigma}_{xx}$ and ${\sigma}_{xy}$ are the longitudinal and Hall components
of the conductivity tensor) is approximately proportional to 
$1/T$ at high temperatures,\cite{cheong} while the cotangent of the Hall angle, 
$\cot{\Theta_H} = {\sigma_{xx}}/{\sigma_{xy}}$ is described by the quadratic
temperature dependence,\cite{chien}

\begin{equation}
\cot{\Theta_H} = \alpha T^2+C, 
\label{cotH}
\end{equation}

\noindent
where $C$ is the impurity contribution. 

The Drude model leads to the same $T$--dependence for $\rho$ and 
$\cot{\Theta_H}$, while $R_H$ should be constant.\cite{ziman}
The orbital magnetoresistance (OMR), 
${\Delta}{\rho}/{\rho}$, should vanish in materials with an
isotropic Fermi surface, and be positive and proportional to 
${\tau}^2$ for an anisotropic Fermi surface in the weak--field regime.\cite{ziman}
A positive OMR has indeed been observed by 
Harris {\em et al.} for two single crystals of YBa$_2$Cu$_3$O$_{7-\delta}$ (YBCO), 
and for an optimally doped single crystal of La$_{1.85}$Sr$_{0.15}$CuO$_{4}$ 
(LSCO).\cite{harris} However, the OMR has been found proportional to $T^{-4}$, 
and cannot 
be described by the same relaxation time as $\rho$. Instead, at high enough
temperatures the OMR follows the same $T$-dependence as ${\tan^2}{\Theta_H}$,

\begin{equation}
{\Delta}{\rho}/{\rho} = {\zeta}{\tan^2}{\Theta_H}.
\label{mrtan}
\end{equation}

The relations (1-2) have been studied in a large variety of high-$T_c$
materials,\cite{gf,takagi,chien,harris2,carr,xiao,kend,ito,wuyts,ando,kons}  
including underdoped, overdoped, and impurity-doped ones.
Various deviations are found in underdoped and overdoped 
LSCO,\cite{takagi,hwang} and YBCO,\cite{carr,ito,wuyts}
and in optimally doped and overdoped single--layer and bilayer
bismuth compounds.\cite{ando,kons} The relation of these  deviations
to the opening of a pseudogap in the normal--state excitation spectrum
is still
under discussion.\cite{takagi,kons,batlogg,timusk} 
Conventional Fermi--liquid behavior is approached with overdoping. 
\cite{takagi,hwang} 

Previous measurements of the magnetoresistance 
are limited and do not include the influence of
impurities.\cite{harris,ando,kimura,tyler,balakir} 
In our earlier study of La$_{2-x}$Sr$_{x}$CuO$_{4}$ films with 
$x$ between 0.048 and
0.275, deviations from Eq.~3 were found both in the underdoped
and the overdoped regime,\cite{balakir} and we suggested a link 
with the pseudogap opening.

The theoretical interpretation of these observations is controversial. 
Fermi--liquid (FL) models assume that strongly anisotropic scattering 
results in ``cold spots'' and  ``hot spots'', small areas on the Fermi surface 
which preferentially contribute to the transport properties and produce 
temperature anomalies.\cite{carr,kend,stoj,hlub,ioffe,zhel1,zhel2} 
A somewhat different proposal is the two-patch model,\cite{PSK} 
where anisotropic scattering results in cold and hot patches, 
sizeable areas of the Fermi surface with different 
properties and different temperature dependences of
the scattering amplitudes, which result in  transport anomalies.
Some of these models lead to nearly $T$-independent results for the
coefficient $\zeta$. Most of them predict an inpurity dependence for $\zeta$.

Other approaches include
non-FL models, which attribute the transport behavior to the
existence of two different relaxation rates at each point on the Fermi surface,
separately governing the longitudinal and transverse (Hall) 
scattering.\cite{anders,col} 
In these models $\zeta$ is independent of temperature, and impurities. 
Measurements of the infrared Hall effect give support to 
non-FL behavior,\cite{cerne} although the details do not seem to agree with any
of the models. Recently, a theoretical approach based on a marginal FL
hypothesis with strongly anisotropic (forward) impurity scattering 
has been proposed.\cite{va} It was motivated by
angularly resolved photoemission spectroscopy (ARPES) studies\cite{valla}
which show that the single--particle scattering rate contains a
momentum--dependent term, constant in temperature, and a momentum--indepedent 
term linear in $T$, as in the marginal FL models.

To study the validity of these different approaches
requires careful comparison with the experimental data, including 
tests of all of the relations 1 to 3. In particular, a simultaneous study of the 
impurity dependence of the magnetoresistance and the Hall effect can
show whether the coefficient $\zeta$ is impurity--dependent, and
provide 
a new and sensitive test of the theories. 

In this paper we present a study of the resistivity, the Hall effect, and
the magnetoresistance of $c$-axis aligned 
La$_{1.85}$Sr$_{0.15}$Cu$_{1-y}$Zn$_y$O$_4$ films, with values of
$y$, from 0 to 0.12, i.e. almost up to the composition at which
the metal--insulator transition occurs at $y=0.14$.\cite{karpik} 
In the course of this study we found that the films grow with built-in 
strain resulting from the lattice mismatch between the substrate
and the film. The strain is relieved partially by dislocations at the
interface, resulting in a variable amount of strain from film to film. 
Substrate--induced compressive in--plane strain is known to enhance the
superconducting transition temperature,\cite{sato1,sato2,locq} and
the origin of this effect is under discussion.\cite{pv2000} The compressive
in--plane strain is accompanied by a decrease of the residual 
resistivity.\cite{sato2}
The influence of strain on the other transport parameters has not been
previously investigated. Careful structural characterization allows us
to use the variable strain
in the films not only to evaluate its influence 
on the normal--state transport properties, but also to separate its
effect out, and so to get a more precise measure of the effect of impurities.

The outline of the paper is as follows. In section II we describe the experimental
details. The study of the influence of strain on the structure, microstructure,
and transport in LSCO films with $y = 0$ (without zinc) is described in 
section III. Section IV contains the description of the normal--state
transport properties of the films with zinc.
A  comparison with other experiments and 
various theoretical models is
given in section V, followed by a summary in section VI. Some of
the results of this study have already been described in a brief
publication.\cite{ciep}

\section {Experiment}

The specimens used in this study were $c$--axis aligned films grown by pulsed 
laser deposition on LaSrAlO$_4$ (LSAO) single-crystalline substrates.\cite{mzc} 
The substrates were oriented with the $c$-axis 
perpendicular to the substrate surface to better then 0.2 deg.
During deposition the substrate was held at 
720 $^{\circ}$C in an oxygen atmosphere of 100 mTorr. The energy density 
of the laser pulse was held to about 1.5 J/cm$^2$, and
the frequency to 2.1 Hz. The growth rate was about 0.25 \AA\ per pulse.
After deposition, the oxygen pressure was increased to 750 Torr, and the
films were slowly cooled to room temperature over a period of 2 hours.
The thickness of the films was in the range from 5000 to 9000 \AA\, and we have 
verified that a change of thickness within this range does not
influence the
superconducting and transport parameters.  

To separate the effects of the impurities
from those of strain and microstructure, two sets of films were made.
One set, with nominal composition La$_{1.85}$Sr$_{0.15}$CuO$_4$ (LSCO)
and $T_c$ between 25 K and 35.2 K, was used to evaluate the correlation
between the structural and the superconducting properties. In the second set of 
films, with nominal composition of 
La$_{1.85}$Sr$_{0.15}$Cu$_{1-y}$Zn$_y$O$_4$ (LSCZNO), the zinc fraction, $y$, 
was varied from 0 to 0.12. X-ray diffraction and absorption measurements 
confirmed that the value of $y$ in the films was close to the nominal
values in the targets.\cite{aplzn} 

The specimens for the transport measurements were patterned by photolithography,
and wires were attached with indium to evaporated silver pads. 
Simultaneous measurements of the Hall effect and the magnetoresistance were
in a standard six--probe geometry in a magnetic field up to 8 T,
perpendicular to the CuO$_{2}$ planes and to the current direction 
(transverse configuration), and for both field orientations.
The magnetoresistance was also measured in the longitudinal configuration, i.e. 
with the magnetic field parallel to the CuO$_{2}$ planes. The temperature 
was varied from 25 K to 300 K, and measured with a Cernox sensor, whose
resistance was
stabilized to about 3 parts per milllion.\cite{balakir} The Hall voltage was
a linear function of the magnetic field up to the highest fields
used in this study.

The $c$--axis lattice parameters were determined from eight high--angle $(00l)$
diffraction peaks in the ${\theta}- 2{\theta}$ scans measured with a
Rigaku X--ray diffractometer. Both K$\alpha_1$ and
K$\alpha_2$ peaks were fitted with Lorenzians, and the parameters
were calculated from the least-square fit to all peak positions. The
rocking curves were measured for the (008) peak. Several specimens
were also studied with a 4--cycle diffractometer to determine the magnitude of the
$a$-axis lattice parameter. 
The topography of the films was evaluated using atomic force microscopy,
with a Park Scientific Instruments AutoProbe M5.

\section {LSCO films: relation between microstructure and transport
properties}

\subsection {Resistivity and Hall effect}

\begin{figure}[ht]
\vspace*{-1.3cm}
\epsfig{file=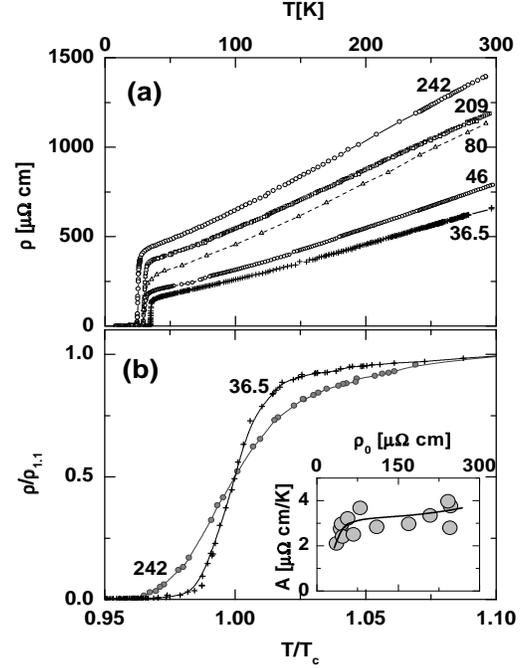, height=0.76\textwidth, width=0.56\textwidth}
\vspace*{-3.2cm}
\caption{(a) The temperature dependence of the resistivity for several
La$_{1.85}$Sr$_{0.15}$CuO$_{4}$ films. 
The data are labeled with the value of $\rho_0$ for each film (in 
${\mu}{\Omega}\,$cm). (b) The resistivity, normalized to the value at 1.1$T_c$, 
$\rho_{1.1}$,
versus the reduced temperature, $T/T_c$, for two films, with the lowest, and the 
highest $\rho_0$. Inset: $A$ versus $\rho_0$ for larger set of LSCO films. The line is
a guide to the eye.}
\label{fig1}
\end{figure}

In spite of identical growth parameters, the specimens grow with various 
resitivities and values of $T_c$. Fig.~1a shows several typical examples of the
temperature dependence of the resistivity. The data are labeled with the value
of the residual resistivity, $\rho_0$, determined from the fit of equation
(\ref{resis}) between 200 K and 300 K. 
\begin{figure}[ht]
\vspace*{-0.8cm}
\epsfig{file=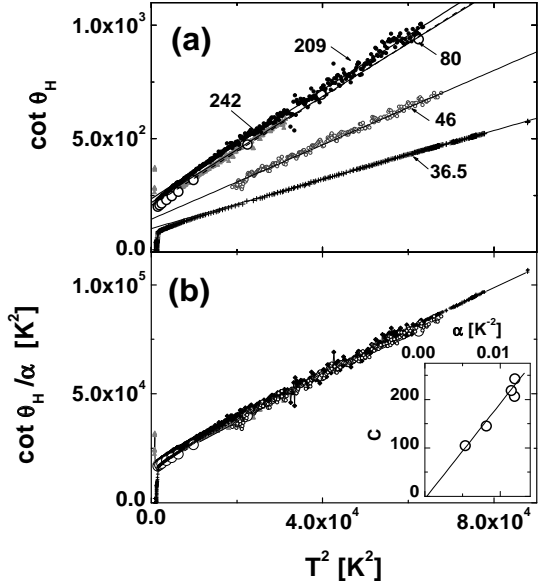, height=0.74\textwidth, width=0.52\textwidth}
\vspace*{-4.0cm}
\caption{(a) $\cot{\Theta_H}$ at 8 T as a function of $T^2$ for the same 
films as in Fig.~1a. The lines are fits to the relation 
$\cot \Theta_H = \alpha T^2 + C$ and the data are labeled with the value 
of $\rho_0$. (b) $\cot{\Theta_H}/\alpha$ as a function of $T^2$: the data
collapse to one line. Inset: $C$ vs. $\alpha$ for the same set of films.}
\label{fig2}
\end{figure}
It is seen that the increase
of $\rho_0$ is accompanied by a decrease of $T_c$, from 35.2 K in the
film with the highest to 25 K in the film with the lowest $T_c$. The details 
of the superconducting transition are shown in Fig.~1b for these two films. 
The lowering of  $T_c$ is associated with a substantial
broadening of the transition. 

Interestingly, the slope of the $T$--dependence of the resistivity, $A$,
does not grow as fast as the residual resistivity. This is shown in
the inset in Fig.~1b where we include the data from a larger set of LSCO films.
While $\rho_0$ increases by a factor of about seven in this set,
$A$ increases by a factor of at most two.

Fig.~2 shows the cotangent of the Hall angle measured in a magnetic
field of 8 tesla as a function of $T^2$ for the same set of films as in Fig.~1a. 
\begin{figure}[ht]
\vspace*{-0.8cm}
\epsfig{file=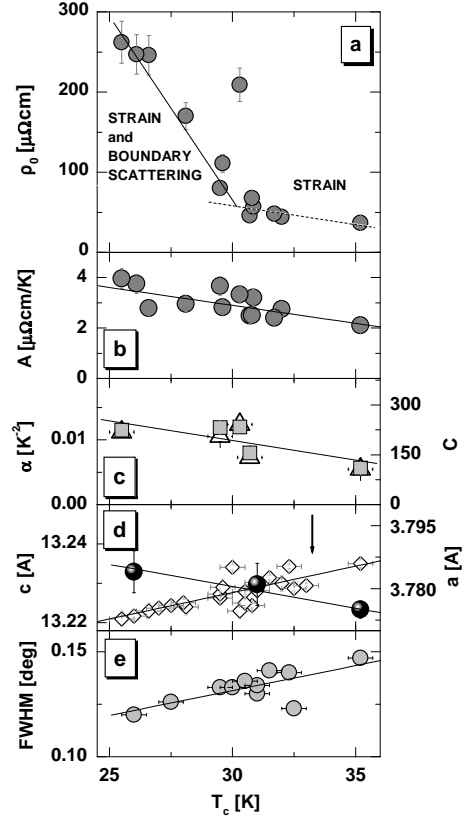, height=0.72\textwidth, width=0.50\textwidth}
\vspace*{-1.0cm}
\caption{The correlation between $T_c$ and several transport and structural 
parameters: (a) $\rho_0$, (b) $A$, (c) $C$ (triangles) and $\alpha$ (squares), 
(d) $c$ (diamonds) and $a$ (circles), (e) FWHM. All
lines are linear fits to the data. The arrow in (d) indicates the bulk
values of the lattice parameters.}
\label{fig3}
\end{figure}
It is seen that the data follow straight lines except at the lowest temperatures.
The values of the intercept $C$ and the slope 
$\alpha$ of the Hall--angle lines from the fit to Eq. 2 are shown
in the inset to Fig.~2b. Remarkably, $C$ is proportional to
$\alpha$, in contrast to the relation between $\rho_0$ and 
$A$ shown in Fig.~1. 
The proportionality between $C$ and $\alpha$ is also illustrated in Fig.~2b
which shows $\cot{\Theta_H}$ divided by $\alpha$. It is seen that
the data for films with different values of $\rho_0$ collapse to a single line.

The correlation between the superconducting transition temperature and the
transport parameters $\rho_0$, $A$, $C$ and $\alpha$ for the LSCO films 
is summarized in Figs.~3a to 3c. We see that as $T_c$ decreases,
all of these 
parameters increase. The fractional changes of $A$, $C$, and $\alpha$ are
closely the same, about 8.3\% per kelvin. $\rho_0$ follows the 
same dependence, with the same fractional changes, in the high--$T_c$ regime, 
between 35 K and 30 K. On the other hand, below 30 K there is a crossover 
to a regime where $\rho_0$ increases faster then any of the other 
parameters. 

An increase of $\rho_0$ may be caused by several effects. A conceivable
origin is a change of chemical composition such as a deficiency
of strontium or oxygen, which could create scattering centers and 
decrease the carrier concentration.
However, this cannot be the sole cause of the observed effects. A decrease in
carrier concentration is known to lead to a decrease of the slope 
of the cotangent line,\cite{hwang,balakir} contrary to the result shown in Fig.~3c. 
This suggests that some other type of disorder is present in the films.
The fact there are
two different regimes with different behavior of $\rho_0$ suggests that there
might be two effects.

Compared to the large 
increase of $\rho_0$, the change of $T_c$ is quite small. When $T_c$ drops
by 5 K (from 30 K to 25 K), $\rho_0$ increases by 200 ${\mu}{\Omega}\,$cm
so that the rate is about $2.5 \times 10^{-2}\,$K/${\mu}{\Omega}\,$cm. 
As we will show later, this rate 
is 4 to 5 times larger when $T_c$ is suppressed by zinc impurities.
That the disorder 
in the films contributes so much less effectively 
to the suppression of $T_c$ than the impurities suggests that 
points defects are not likely to be responsible.

\subsection {Dislocations and strain}

There are  several possible origins of disorder in the LSCO films.
They include strain resulting from the lattice mismatch, imperfections
in the microstructure (including dislocations), and changes
in the chemical composition such as strontium or oxygen defficiency.

The in-plane lattice parameters of the film and the substrate differ by 
about 0.5\% (the $a$--axis parameter is 3.756 {\AA} in LSAO and 3.777
{\AA} in LSCO), and compressive in--plane strain should therefore be
expected. It has been observed,\cite{sato1,sato2,locq} and
is accompanied by $c$--axis expansion and by an
increase of the superconducting transition temperature. 

The degree to which the films are strained depends
on the way in which the strain is relieved. If the films are very
thin, of the order of several unit cells, the lattice constant may remain 
close to that of the substrate throughout the 
film.  For thicker films, however, the strain is usually relieved
to some degree. This may occur gradually as the distance from the
substrate increases, or by
dislocations right at the surface.\cite{disloc} The dislocations, in
turn, may contribute to other structural imperfections.

Even with the same substrate, and the same lattice--film mismatch,
the microstructure may be different depending on the deposition rate and
other deposition parameters. It has been reported, for example, that
reactive co--evaporation results in layer--by--layer
growth of SrTiO$_3$ films for small deposition rates, while a more
disordered 3D--type of growth is observed for larger rates.\cite{satnai} 
Pulsed laser deposition puts a large amount of 
material very rapidly on the substrate surface. Under these conditions 
tiny fluctuations in the growth parameters may have a substantial effect.
Even though we attempt to keep all growth parameters constant, some
variations in the film microstructure is difficult to avoid. 

In order to identify the possible causes for the disorder and for the 
change in the transport parameters, we have made a careful
study of the structure and microstructure of the LSCO films. The details
of this study will be reported elsewhere.\cite{mcstr} Here we
list some results which have implications for the
transport and superconducting properties. 

{\em (1)} X--ray diffraction (${\theta}-2{\theta}$ scans) indicates
that the $c$--axis
lattice parameter decreases linearly with the decrease of $T_c$, as
shown in Fig.~3d. This indicates that films grow with various degrees
of strain. The error bars are standard deviations
from the average $c$--value calculated from eight high--angle $(00l)$ peaks.
In the samples which are shown in the figure the standard deviation is very small
(less then 0.001 {\AA}). A small number of films (not shown) exhibit large
standard deviations, indicative of a real distribution of $c$--values.
We interpret these as the films with gradual strain relief. Conversly,
in the specimens shown in Fig.~3d this gradual change of $c$ is absent,
and strain relief presumably occurs by dislocations at the interface.

{\em (2)} $\phi$--scans obtained with a  4--cycle diffractometer
indicate fourfold in-plane   
symmetry which persists in specimens with low $T_c$, indicating that
the good
in--plane alignment is not disturbed by disorder. The $a$--axis parameter
increases with a decrease of $c$ and $T_c$ (Fig.~3d) as one should expect 
for strain induced by lattice mismatch. The arrows in Fig.~3d show the 
positions of the bulk lattice parameters. The in--plane strain changes from 
compressive (high $T_c$), to negligible ($T_c \approx 30-32$ K), and
then to tensile 
($T_c < 30$ K). We see that the large disorder that gives rise to the enhanced 
$\rho_0$ is associated with tensile in--plane strain.

{\em (3)} In the films with large $c$--distributions the full widths at
half maximum (FWHM) of the rocking curves are quite large,
as high as 0.4 deg in some cases (not shown on the figures),
supporting our previous conclusion that strain relief takes place
gradually in these films. 
On the other hand, FWHM remains small in the majority of specimens (Fig.~3e),
confirming the good crystalline quality of the films. In fact, FWHM becomes smaller
as $T_c$ drops, indicating that crystalline quality, at least as seen
by X--ray measurements, improves  with tensile strain. This suggests
that random disorder, as would result from oxygen or strontium deficiency, is
not present, or at least does not increase as $T_c$ decreases.

\begin{figure}[ht]
\vspace*{-1.5cm}
\epsfig{file=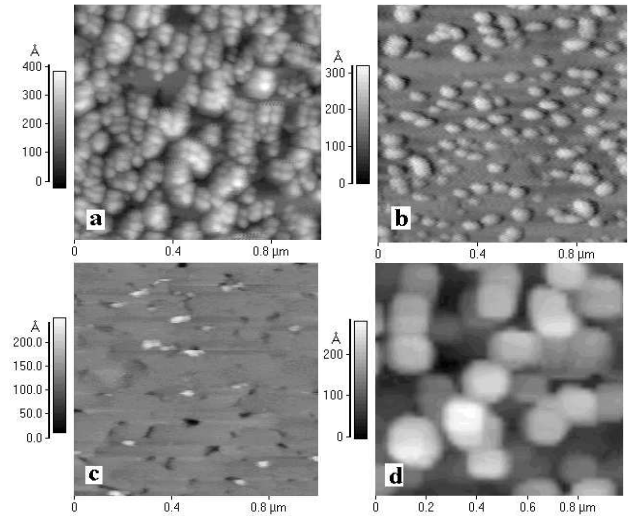, height=0.80\textwidth, width=0.60\textwidth}
\vspace*{-5.3cm}
\caption{AFM images of LSCO films with various values of $T_c$ and FWHM: 
(a) 32 K and 0.41 deg, (b) 35.2 K and 0.147 deg, (c) 31 K and 0.134 deg,
(d) 26 K and 0.12 deg.}
\label{fig4}
\end{figure}

{\em (4)} A direct look at the microstructure of the 
films can be obtained with Atomic Force Microscopy (AFM)(Fig.~4). We
find that the films with gradual strain relief grow with grains of 
various sizes, from  
about 50 nm to 200--500nm (image a), 
characteristic for 3D growth. The root--mean--square (rms) roughness 
is as large as 7 nm in these films. 
The grains in the films without this gradual strain relief are more uniform
in size (images b to d). The cross--sections of the images indicate
that the films start
as islands, several unit cells high at a time. In the films with
compressive or negligible strain these islands coalesce into quite smooth surfaces,
with rms roughness of the order of one to two unit cells. The images b and c
represent this regime, with the difference between them resulting from different
times at which the growth terminates.
When the strain becomes more and more tensile substantial imperfections appear 
at the grain boundaries when the grains coalesce. In the film with the lowest 
$T_c$ we observe the formation of very flat, imperfecty connected grains (image d). 
The grains are quite uniform in size, about 150 {\AA} across, with very 
good crystalline quality inside the grains. The rms roughness 
increases again, to about 6 nm. In the tensile regime the appearance of
imperfect grain boundaries  
is well correlated with the enhancement of the residual resistivity,
which we therefore associate with grain boundary scattering.

We interpret the evolution of the microstructure as resulting from the
the increase of the number of dislocations created at the substrate--film interface.
For compressive or negligible strain, a small number of dislocations
relieves the strain partially or fully, contributing to a change of strain from
film to film. $T_c$ remains high, but changes with strain. As the
number of dislocations increases, they are built in at the grain boundaries,
contributing to the tensile strain in the grains, to enhanced scattering, 
and to a decrease of $T_c$. We can conclude that there are two
regimes in the transport properties of the LSCO films. In one, where $T_c$ is larger
than about 28 or 30 K, and $\rho_0$ is smaller than about $100\,{\mu}{\Omega}\,$cm,
the change of $\rho_0$ and $T_c$ is a result of the change of strain alone.
In the second the transport properties and $T_c$ are affected by 
grain--boundary scattering. These two regimes are shown on Fig.~3a as
``strain'' and ``strain and boundary scattering''.

\begin{figure}[ht]
\vspace*{-1.5cm}
\epsfig{file=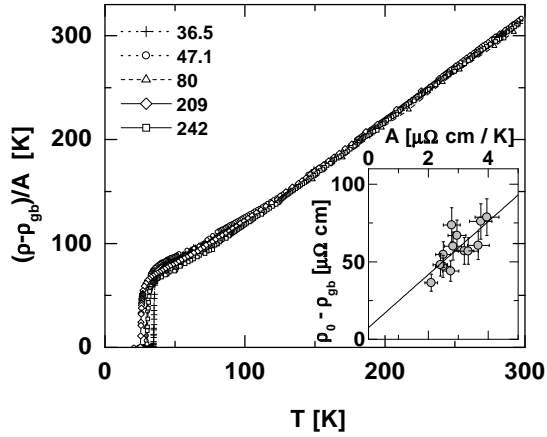, height=0.66\textwidth, width=0.46\textwidth}
\vspace*{-4.0cm}
\caption{$(\rho - \rho_{gb})/A$ as a function of $T$, where $\rho_{gb}$ is the
grain boundary resistivity. The samples are the
same as on Fig.~1a. The data are labeled with the value of $\rho_0$ 
(in ${\mu}{\Omega}\,$cm) for each film. 
Inset: $\rho_0 - \rho_{gb}$ as a function of $A$
for a larger set of films. The line is fitted to the data.}
\label{fig5}
\end{figure}

The fact that $A$, $C$, $\alpha$, and in the strain regime also $\rho_0$,
undergo the same fractional changes, suggests that the strain affects
all transport coefficients in the same 
way, while grain--boundary scattering affects only the
residual resistivity, introducing an additive contribution to $\rho_0$, 
which we call $\rho_{gb}$.
If this is indeed the case then the quantity $\rho_0 - \rho_{gb}$ should
depend on strain only, and therefore should be proportional to $A$. 
To estimate this quantity we first extrapolate the linear relation between $\rho_0$ and 
$T_c$ from the strain regime, where $\rho_{gb} = 0$, as shown by 
the dashed line in Fig.~3a, and
calculate the value of $\rho_0 - \rho_{gb}$ for each sample from its
value of $T_c$. The inset to Fig.~5 shows that the calculated
$\rho_0 - \rho_{gb}$ is indeed closely proportional to $A$.
The temperature dependence 
of $\rho - \rho_{gb}$ divided by $A$ is shown in the main part of Fig.~5 
for the same set of films as displayed in Fig.~1. 
We see that the data for different
films collapse on the same line, with small deviations appearing only below
100 K, probably because of imperfect estimates of $\rho_{gb}$.
This result confirms the consistency of our 
analysis and shows that the conclusions about the distinct effects of strain
and grain--boundary scattering are correct.

From the fit to the data in Fig.~3d we estimate that the
rate of decrease of $T_c$ with $c$ is equal to 680 K/${\AA}$.
This value is more than three times as large as the rate reported by 
Sato {\em et al.}\cite{sato2} for LSCO films with
$0.12 < x < 0.18$.
It is possible that a variation in the amount of strontium affects the rate, or 
that the microstructure of films prepared by reactive coevaporation\cite{sato2}
is different from that of films grown by pulsed laser deposition, so that
a straighforward comparison of the results is difficult. 

\begin{figure}[ht]
\vspace*{-0.8cm}
\epsfig{file=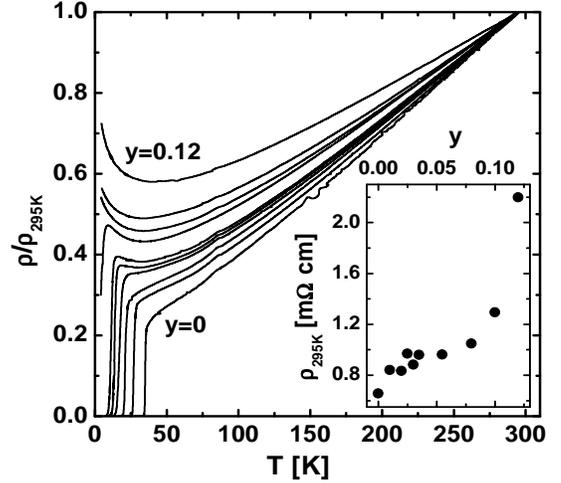, height=0.75\textwidth, width=0.55\textwidth}
\vspace*{-6.0cm}
\caption{Temperature dependence of the resistivity (normalized to room-temperature) 
of a series of La$_{1.85}$Sr$_{0.15}$Cu$_{1-y}$Zn$_y$O$_{4}$ films 
with $y = 0, 0.01, 0.02, 0.025, 0.03, 0.035, 0.055, 0.08, 0.10, 0.12$ (from bottom to top).
Inset: the room-temperature resistivity as a function of $y$.}
\label{fig6}
\end{figure}

Defining the strain as $\epsilon_d =(d_{bulk}-d_{film})/d_{bulk}$ (where $d$ is the
lattice parameter), we see that in our films it ranges from about 
$\epsilon_a = +0.05$\% and $\epsilon_c = -0.02$\% (compressive in--plane strain) 
for the film with the highest $T_c$, to about $\epsilon_a = -0.19$\% and $\epsilon_c =
+0.08$\% (tensile in--plane strain) for the film with the lowest $T_c$.
The change of $T_c$ induced by the compressive or tensile strain may be found
from the expression,\cite{locq} 
$T_c=T_c (0)+2({\delta T_c}/{\delta \epsilon_{ab}})\epsilon_{ab}
+({\delta T_c}/{\delta \epsilon_{c}})\epsilon_{c}$, where 
$\delta T_c/\delta \epsilon _{i}$ ($i = a,b$, or $c$) are the uniaxial strain 
coefficients. The values of these coefficients were estimated in Ref.[45], 
$\delta T_c/\delta \epsilon _{a} = (250 \pm 340)\,$K, 
$\delta T_c/\delta \epsilon _{b} = (400 \pm 340)\,$K,
and $\delta T_c/\delta \epsilon _c = (-1090 \pm 1120)\,$K.
Using the average of the values for the $a$ and $b$ axes as $325\,$K, and for
the $c$--axis as $-1090\,$K, we estimate that compressive and tensile strain
should change $T_c$ by about $+1\,$K and $-2\,$K, respectively,
so that the maximum change of $T_c$ directly related to strain should
not exceed $3\,$K, while
the experiment shows a change of the order of $10\,$K. With the
large uncertainty of the strain coefficients the agreement may not be
unreasonable. There might be some additional contribution to the suppression
of $T_c$ in the tensile--strain regime with grain--boundary scattering 
from partial isolation of the grains and weak links between them.
This would also explain the broadening shown in Fig.~1b.\cite{link} 

\section {LSCZNO films: impurities and strain}
\subsection {Resistivity}

We now turn to the specimens in which zinc is substituted
for some of the copper. The temperature dependence of the resistivity
for some of these films (normalized to the value  
at room temperature, $\rho_{295K}$), is shown in Fig.~6.
The inset shows the dependence of $\rho_{295K}$ on $y$.
These values are about 30\% higher than in similar single crystals.\cite{fukuz} 
However, while in the single crystals zinc can be added only up to
$y=0.04$, three times as much can be added in the films.
The room--temperature resistivity increases
linearly with $y$ up to about
$y = 0.1$. For larger $y$ the increase is faster than linear as a result
of the approach to the metal--insulator transition, as described
in a previous publication.\cite{karpik} The films remain superconducting
up to $y = 0.055$, and the films with $0.055 < y < 0.12$ are metallic but 
nonsuperconducting. In the ceramic specimens superconductivity dissapears for 
$y = 0.03$. \cite{marta3} 

In all films the resistivity is linear in $T$ at high temperatures. 
We fit the high--$T$ resistivity
between 200 K and 300 K with Eq.~1 to obtain the residual 
resistivity. Just as in the
case of undoped LSCO, the films grow with various values of
$T_c$ and residual resistivity for any given value of $y$. The
correlation between $T_c$ and 
$\rho_0$ for films with zinc was already discussed
in Refs.[41,34], where we found that for films with small $\rho_0$ the
dependence of ${T_c}$ on $\rho_0$ is well described by the
Abrikosov--Gorkov formula.\cite{ag}  
On Fig.~7 we show $\rho_0$ as a function of $T_c$. 
We include 
the data for $y = 0$ of Fig.~3a as well as the fit to the
Abrikosov--Gorkov formula from Ref.[34]. We see that the crossover
in the $\rho_0$--vs--$T_c$ line, observed for the films with $y = 0$, is also present
for other values of $y$, so that we expect our interpretation of
strain, dislocations, and grain boundary scattering to be the same in
the films with zinc as in those without.

\begin{figure}[h]
\vspace*{0.0cm}
\epsfig{file=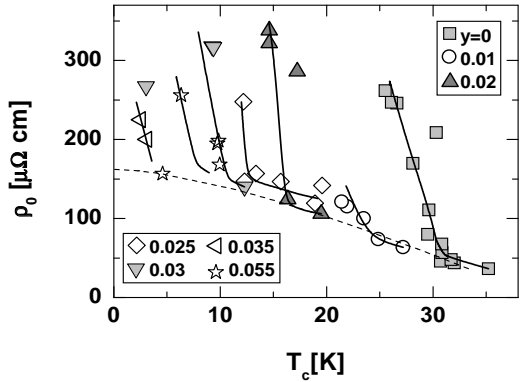, height=0.70\textwidth, width=0.50\textwidth}
\vspace*{-5.8cm}
\caption{The residual resistivity $\rho_0$ as a function of $T_c$ for
films with various values of $y$. The dashed line
is drawn through the data for films with $\rho_s < 100 {\mu}{\Omega}\,$cm,
which follow the Abrikosov-Gorkov formula. All other lines are guides to the
eye.}
\label{fig7}
\end{figure}

\begin{figure}[h]
\vspace*{-1.2cm}
\epsfig{file=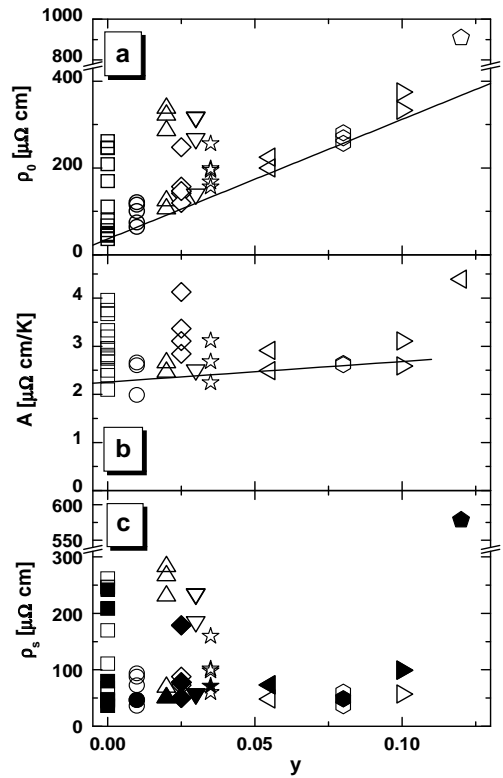, height=0.70\textwidth, width=0.50\textwidth}
\vspace*{-1.0cm}
\caption{The residual resistivity $\rho_0$ (a), the resistivity slope $A$ (b),
and strain-related residual resistivity, $\rho_s$ (c), as a function of $y$. 
The black points in (c) indicate films for which Hall effect was measured. }
\label{fig8}
\end{figure}

We now proceed to the separation of
the effects of strain and grain--boundary scattering from the effect of the
zinc impurities. On Fig.~8a we show the dependence of $\rho_0$ on $y$, for
all the films with various values of the residual resistivity for each
value of $y$. We see that the minimum value 
of $\rho_0$ increases linearly with increasing $y$. We assume 
that this minimum value is the residual resistivity associated with
the zinc impurities, and call it $\rho_y$. The rate of increase of $\rho_y$
with $y$ is $\rho_y /y = 2.8\,$m${\Omega}\,$cm. This value is comparable
to the rate observed in single crystals of YBCO with zinc.\cite{chien} 
(Because of its closeness to the metal--insulator transition we have
excluded the film with $y = 0.12$ in the analysis.)

The slope, $A$, is shown as a function of $y$ in Fig.~8b.
In contrast to $\rho_0$ it does not change dramatically with $y$.
In fact, if we fit a straight line to the data for the films with the minimum
residual resistivity for each $y$ (again excluding $y = 0.12$), we find that $A$
changes by about 17\% as  $y$ changes from zero to $y = 0.1$, not far
from the
experimental error of about $\pm 10$\%.
The same conclusion of a very weak influence of zinc impurities on $A$
may be reached from data for single crystals of YBCO and LSCO,\cite{fukuz}  
although an earlier report noted a substantial increase of $A$.\cite{chien} 
The variability of the results may
be related to differences in the microstructure of the specimens rather
than to a change of $A$ with $y$. In any case, we conclude that 
$A$ remains almost unchanged by the addition of the impurities.
 
\begin{figure}[h]
\vspace*{-1.0cm}
\epsfig{file=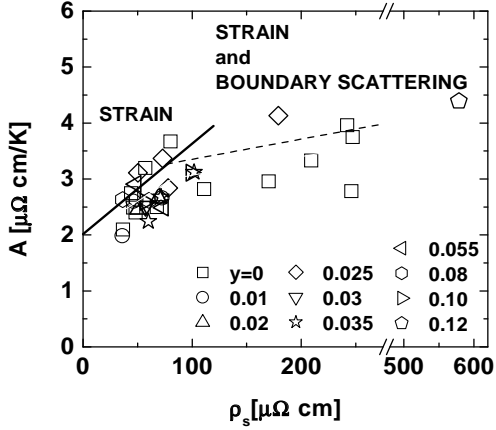, height=0.65\textwidth, width=0.46\textwidth}
\vspace*{-5.0cm}
\caption{$A$ as a function of $\rho_s$ for films with different values
of $y$. The solid line is a linear fit in the strain regime and the dashed line is
a fit for films with grain--boundary scattering, with $\rho_s$ larger than
about 100 ${\mu}{\Omega}\,$cm.}
\label{fig9}
\end{figure}

If $\rho_y$ is the zinc-related part of the residual resistivity, then the
difference, $\rho_s = \rho_0 - \rho_y$, is the part which results from
strain and grain--boundary scattering.
Its dependence on $y$ is shown on Fig.~8c, where we see that its
minimum value does not depend on $y$.

The separation into a $y$--dependent and a $y$--independent part shows that
Matthiesson's rule is valid and that we can have confidence in the relation

\begin{equation}
\rho_0 = \rho_y + \rho_s .
\label{rhos}
\end{equation}

We can test this relation further by investigating 
the relation of $\rho_s$ to the other transport parameters.
The relation of $A$ to $\rho_s$ is shown on Fig.~9. In spite of
considerable scatter we can identify two different 
regimes. When $\rho_s$ is smaller
then about $100\,{\mu}{\Omega}\,$cm, $A$ increases rapidly, while for
larger $\rho_s$ a crossover is seen to a much weaker increase. 
This crossover is similar to that of Fig.~3a, i.e. it is
caused by more rapid increase of $\rho_s$ in the 
grain--boundary scattering regime. 
We will see that a similar conclusion can be drawn from the
Hall--effect data. 

\subsection {Hall effect}

\begin{table}
\caption{Transport parameters of the films of 
La$_{1.85}$Sr$_{0.15}$Cu$_{1-y}$Zn$_y$O$_4$. The definitions of all parameters
are given in the text. The experimental accuracy is about $\pm$ 15\% for 
the last column, and about $\pm$ 10\% for all other values.}
\begin{tabular}{|c|c|c|c|c|c|c|}
$y$ & $\rho_0$ & $T_c$ & $A$ & $\alpha$ & $C$ & 
${\zeta}(\frac{\alpha_0}{\alpha})^2 $ \\
  & (${\mu}{\Omega}\,$cm) & (K) & ($\frac{{\mu}{\Omega}\,cm}{K}$) & 10$^{-3}$K$^{-2}$ & & \\
\hline
0     & 36.5 & 35.2 & 2.10 & 5.65 & 106 & 11.05 \\
\hline
0     & 46.0 & 30.7 & 2.49 & 8.00 & 142 & - \\
\hline
0     & 80.0 & 29.5 & 3.67 & 11.9 & 206 & 11.15 \\
\hline
0     & 209 & 30.3 & 3.33 & 12.0 & 242 & - \\
\hline
0     & 242 & 25.5 & 3.96 & 11.7 & 218 & - \\
\hline
0.01  & 73.8 & 26.5 & 2.60 & 7.52 & 150 & 10.6 \\
\hline
0.02  & 106 & 21.0 & 2.46 & 7.81 & 166 & 8.00 \\
\hline
0.025 & 119 & 20.4 & 3.11 & 8.22 & 194 & 7.21 \\
\hline
0.025 & 147 & 17.1 & 2.84 & 8.85 & 220 & 7.66 \\
\hline
0.025 & 248 & 13.5 & 4.13 & 13.7 & 340 & 7.01 \\
\hline
0.03  & 140 & 13.6 & 2.51 & 7.47 & 195 & 6.04 \\
\hline
0.035 & 168 & 11.5 & 2.68 & 8.35 & 228 & 6.52 \\
\hline
0.055 & 225 & 3.6 & 3.11 & 10.1 & 323 & 5.11 \\
\hline
0.08  & 269 & 0 & 2.64 & 8.00 & 313 & 3.28 \\
\hline
0.10  & 375 & 0 & 3.11 & 14.6 & 652 & 3.78 \\
\hline
0.12  & 910 & 0 & 4.39 & 21.4 & 1090 & - \\
\end{tabular}
\label{table2}
\end{table}
\begin{figure}
\vspace*{-0.8cm}
\epsfig{file=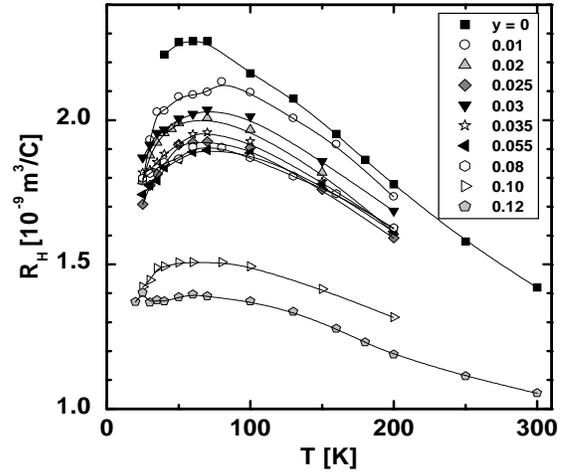, height=0.75\textwidth, width=0.55\textwidth}
\vspace*{-5.5cm}
\caption{Temperature dependence of the Hall coefficient, $R_H$, for a series
of films with various values of $y$.}
\label{fig10}
\end{figure}

On Fig.~8c we show by black points the specimens
for which the Hall effect was studied. The influence of strain on the Hall angle
was already discussed for $y = 0$. To see how the zinc impurities change the
strain--related effects, we have measured several films with $y = 0.025$ and
with different values of $\rho_s$, including the film with 
$\rho_s > 100 {\mu}{\Omega}\,$cm, in the grain--boundary scattering regime.
The transport parameters for the films in which the Hall effect was studied are 
collected in Table 1. The table also includes the data on
magnetoresistance which will be discussed in the next section.

\begin{figure}[ht]
\vspace*{-1.0cm}
\epsfig{file=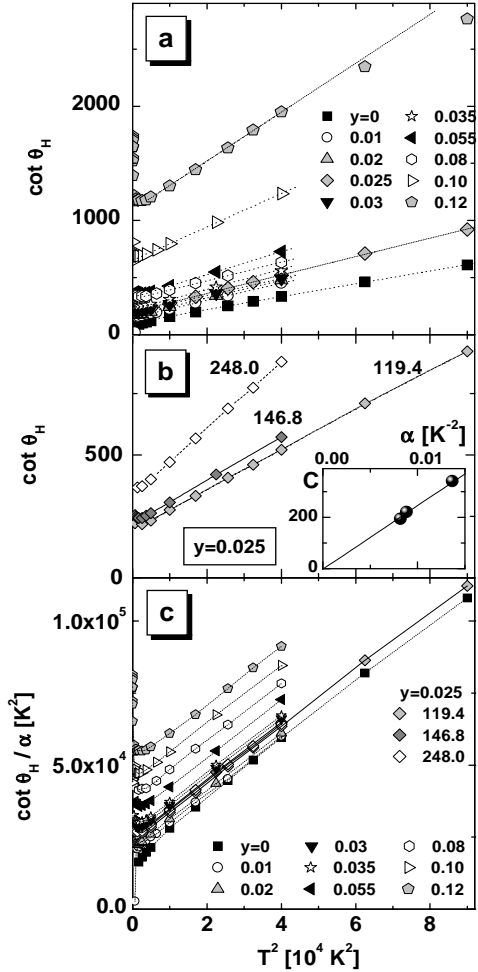, height=0.82\textwidth, width=0.59\textwidth}
\vspace*{-1.0cm}
\caption{(a) and (b) $\cot{\Theta_H}$ at 8 T, (c) $\cot{\Theta_H}/\alpha$ 
as a function of $T^2$. The lines are fits to the equation 
$\cot \Theta_H = \alpha T^2 + C$. In (a) the data are for the films with different 
values of $y$; in (b) the data are for $y = 0.025$ and different values 
of $\rho_0$ (in ${\mu \Omega}\,$cm); in (c) all the data are included. 
The solid lines are drawn through the data for the films with $y = 0.025$. 
Inset in (b): the correlation between $C$ 
and $\alpha$ for films with different values of $\rho_0$.}
\label{fig11}
\end{figure}

On Fig.~10 we show the temperature dependence
of the Hall coefficient for selected films with various values of $y$.
We see that the zinc causes a decrease of $R_H$ without appreciably affecting the
shape of $R_H (T)$. This behavior is quite different from that observed
in LSCO when the strontium--lanthanum ratio is changed, both in the underdoped
and in the overdoped regimes.\cite{hwang,balakir} 
With decreasing
strontium the metal--insulator transition is approached as a result of the
decrease in the carrier concentration. This leads to an
increase of the Hall coefficient by a factor of about
six.\cite{balakir} In the films with zinc, on the other hand, the
decrease of $R_H$ is quite modest,
only by a factor of about 1.6.
This underscores the fact that with the addition of zinc the metal--insulator
transition is quite different, driven by disorder, rather than by a
change in the carrier 
concentration.\cite{karpik} 

\begin{figure}[ht]
\vspace*{-1.0cm}
\epsfig{file=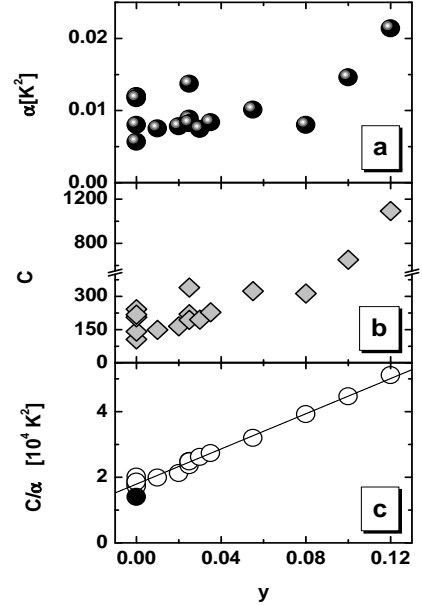, height=0.6\textwidth, width=0.45\textwidth}
\vspace*{-0.5cm}
\caption{$\alpha$ (a), $C$ (b) and $C/\alpha$ (c) as a function of $y$.
The full point in (c) is for La$_{1.83}$Sr$_{0.17}$CuO$_{4}$ single crystal 
of Harris {\em et al.}, Ref.[8].}
\label{fig12}
\end{figure}

The influence of the zinc on the cotangent of the Hall angle is shown
in Fig.~11. Part a shows the data as a function of $T^2$ for films 
with different $y$. 
A few of the specimens were measured to 300 K, the rest to 200 K. It is
seen that 
the data for most of the films follow straight lines, except for upturns
below about 70 K in the films with $y > 0$. The upturns may be related to 
localization effects, or to the opening of a pseudogap.
There is also small deviation
from the straight line in the film with $y = 0.12$ for $T > 200 K$,
presumably because of the  vicinity of the metal--insulator transition.
Apart from these deviations, Eq. 2 is followed, even in 
films with so much zinc that there is no longer any superconductivity.
This is quite different from the behavior observed when the strontium content
is altered in LSCO, when deviations are observed away from optimal
doping, particularly in the overdoped region.\cite{hwang,balakir} 

\begin{figure}[ht]
\vspace*{-1.0cm}
\epsfig{file=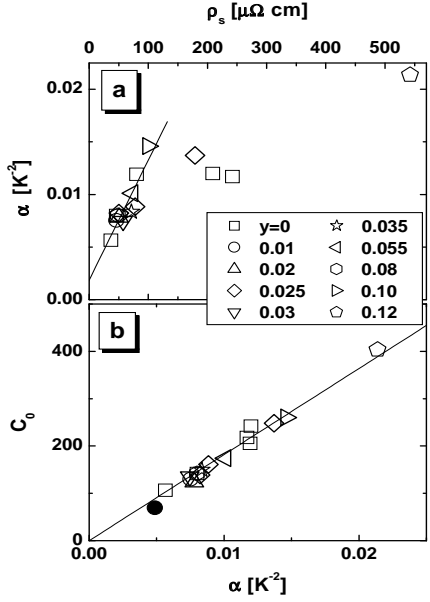, height=0.65\textwidth, width=0.50\textwidth}
\vspace*{-1.5cm}
\caption{(a) $\alpha$ as a function of $\rho_s$ for films with different
values of $y$.
The line is a linear fit in the regime of small $\rho_s$, i.e. without
grain--boundary scattering.
(b) The proportionality bewteen $C_0$ and $\alpha$. The full point is
for the La$_{1.83}$Sr$_{0.17}$CuO$_{4}$ single crystal of Harris {\em et
al.}, Ref.[8].}
\label{fig13}
\end{figure}

An increase of zinc causes an increase in 
the cotangent of the Hall angle. However, the lines do not progress
monotonically 
with $y$. For example, the line for $y = 0.08$ is
below that for $y = 0.055$. The reason is the random variation of strain from
film to film. This can best be seen on Fig.~11b, where we show data for
three films with $y = 0.025$ and different values of the residual
resistivity. The inset shows that $C$ and $\alpha$ are again
proportional to each other. Just as before for $y$ = 0 (Fig.~2), the
three lines 
collapse to one if instead of $\cot{\Theta_H}$
we plot $\cot{\Theta_H}/{\alpha}$ (Fig.~11c). 
Moreover, on this graph the increase of $\cot{\Theta_H}/{\alpha}$ with
increasing $y$ is monotonic. 

This is also illustrated on Fig.~12, where the two top panels 
show $\alpha$ and $C$ as functions of $y$ for all films, including the
three films with $y=0.025$ and the five films with 
$y = 0$ from Fig.~2. The scatter of the parameters disappears when we
plot the ratio $C/\alpha$ (Fig.~12c). The
strain affects both parameters in the same way so that $C/\alpha$ 
is a function of $y$ only. We see further that $C/\alpha$ increases
linearly with increasing $y$.

Fig.~13a shows the dependence of $\alpha$ on $\rho_s$ for
all films measured in this study. For $\rho_s$ less than 100 ${\mu
\Omega}\,$cm the points fall on a straight line, regardless of the
value of $y$. For larger $\rho_s$ there is 
a crossover to a different regime. This is the same
behavior as that of $A$ shown on Fig.~9. It follows that $\alpha$
is $y$-independent, and depends on strain in a similar fashion for all
films regardless of $y$. The relation between $\alpha$ and $\rho_s$ 
is also a further indication that the separation of $\rho_0$ into $\rho_y$
and $\rho_s$ is correct. A previous study of YBCO  crystals with
zinc\cite{chien} without strain finds $\alpha$
also to be $y$--independent, and equal to $5.11 \times 10^{-3}$
K$^{-2}$, close to the minimum in our films.

Since $\alpha$ is $y$-independent, and $C/\alpha$ depends on $y$ only, we
conclude that $C$ is a linear function of $y$: $C = C_0 + C_1 y$. The fit
to the data in Fig.~12c gives two parameters: 
$C_0 /\alpha = (1.8 \pm 0.1) \times 10^4$ K$^2$, and 
$C_1 /\alpha = (2.6 \pm 0.2) \times 10^5$ K$^2$. 

We summarize the behavior of the Hall angle by rewriting Eq. 2 as

\begin{eqnarray}
\cot {\Theta _{H}}/\alpha &=&T^{2}+C/\alpha, \\ 
C&=&C_{0}+C_{1}y.
\label{cotHsc}
\end{eqnarray}

\noindent
where each term in (5) is strain--independent and a function of $y$
only. $\alpha$ and $C$ depend on 
strain in the same way, and $\alpha$ is $y$-independent.

Fig.~13b shows $C_0$ as a function of $\alpha$ for all films. The data
lie closely on a straight line, including even the specimen with $y$ =
0.12, close to the metal--insulator transition, as well as the point
for the single crystal of LSCO\cite{harris}.
 
\subsection {Magnetoresistance}

\begin{figure}[ht]
\vspace*{-0.8cm}
\epsfig{file=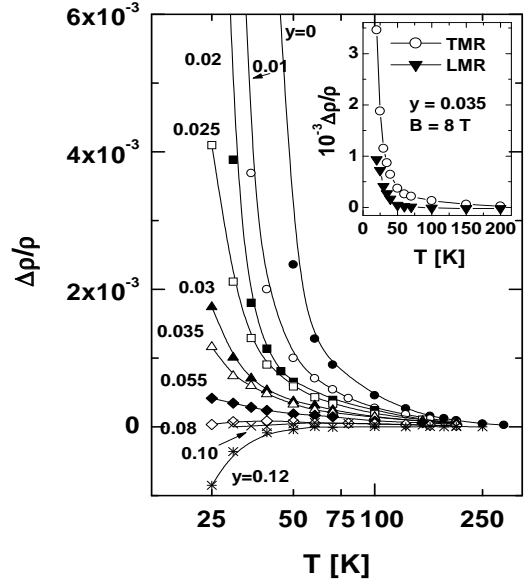, height=0.65\textwidth, width=0.50\textwidth}
\vspace*{-2.5cm}
\caption{The orbital magnetoresistance at 8 tesla as a function of
$\log(T)$
for a series of films with various values of $y$.
Inset: The  TMR and the LMR as a function of temperature for the film with
$y = 0.035$.}
\label{fig14}
\end{figure}

\begin{figure}[ht]
\vspace*{-0.8cm}
\epsfig{file=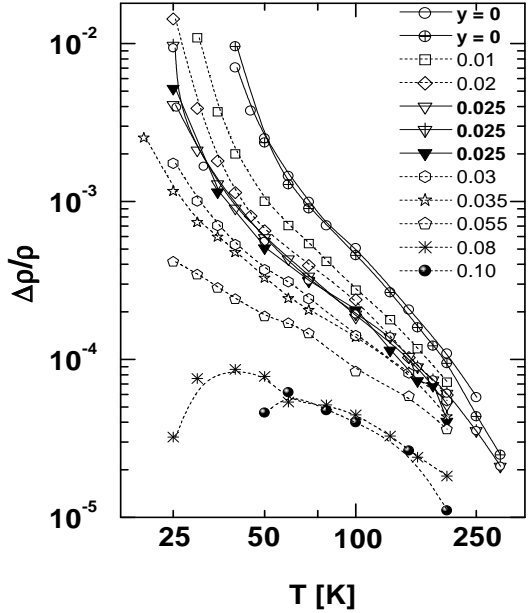, height=0.66\textwidth, width=0.49\textwidth}
\vspace*{-2.5cm}
\caption{Log-log plot of the orbital magnetoresistance at 8 tesla
as a function of temperature
for a series of films with various values of $y$. All lines are guides to
the eye. The solid lines are drawn through the data for $y = 0$ and
$y = 0.025$,
for which several films with different $\rho_0$ were measured.}
\label{fig15}
\end{figure}

The magnetoresistance measurements were performed in two configurations of
the magnetic field with respect to the CuO$_2$ planes. In nonsuperconducting
films, i.e. for $y > 0.055$, both the longitudinal and the transverse 
magnetoresistance (LMR and TMR, respectively) are negative at very low 
temperatures, with increasing magnitude as $y$ increases.
The analysis of this effect, to be presented elsewhere,\cite{artur}  
suggests that it originates in the spin-disorder scattering of 
carriers on magnetically--ordered spin droplets around Zn--impurities. 

In this paper we concentrate on the regime of high temperatures, from 25 K to
300 K. The inset to Fig.~14 shows the $T$--dependences of 
the LMR and the TMR for a film with $y = 0.035$. The TMR is positive over
the whole temperature range, for all values of $y$, and decreases with 
increasing $y$. The LMR has a more complicated behavior, but it is always smaller 
than the TMR, by a factor of 3 to 10, approaching the experimental 
resolution of the measurements at high temperature. At the highest
temperatures the LMR is slightly negative and becomes positive as $T$
is lowered below about 50 to 150 K. 
We attribute the LMR to isotropic spin scattering. Spin scattering
may be expected to grow with increasing $y$ since zinc--doping produces enhanced
staggered magnetization around impurity sites.\cite{julien1}
Subtracting the LMR
from the TMR we obtain the orbital magnetoresistance (OMR), which is shown in the main part
of Fig.~14 for a series of films with various values of  $y$. 

The OMR is positive in most of the films, but decreases rapidly with increasing 
$y$ until, in the film closest to the metal--insulator transition ($y =
0.12$), it becomes negative below about 50 K. 
In fact, we can see
in Fig.~6 that the resistivity of this film shows an upturn in the
vicinity of 50 K. Presumably the negative OMR is related to the localization
effects which dominate the behavior close to the metal--insulator transition. 

In Fig.~15 we show the $T$--dependence of the OMR on a double logarithmic plot
for the films with positive OMR. There are two important features in this
graph. First, the magnitude of the OMR decreases monotonically with increasing
$y$. This suggests that the OMR is not sensitive to strain effects.
This is confirmed by the lines for $y = 0$ and for 
$y = 0.025$, which include data for several films with different 
residual resistivities. They can be seen to be following closely along single
lines for each value of $y$, showing that the
OMR is not influenced by strain but depends solely on $y$.

\begin{figure}[ht]
\vspace*{-0.5cm}
\epsfig{file=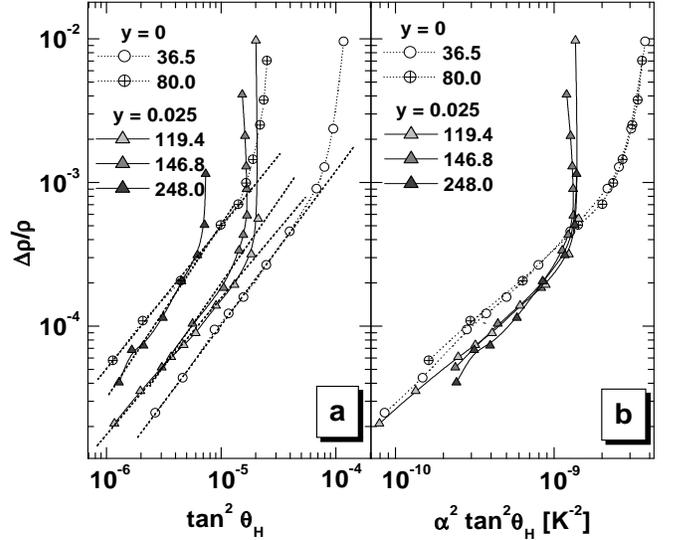, height=0.70\textwidth, width=0.50\textwidth}
\vspace*{-4.0cm}
\caption{The OMR as a function of $\tan^2{\Theta_H}$ (a), and as a
function
of $\alpha^2\tan^2{\Theta_H}$ (b). The data include two films with $y=0$
(circles and dotted lines), and three films with $y=0.025$ (triangles
and solid lines),
with different residual resistivities (in ${\mu \Omega}\,$cm). The dotted
and solid lines
are guides to the eye. Straight dashed
lines are fitted to the data in the high--temperature regions.}
\label{fig16}
\end{figure}

The second interesting feature is the shape of the OMR curves.
The lines for all superconducting specimens ($y > 0.055$) have a
characteristic ``S''--shape, as previously observed in 
optimally doped LSCO and  YBCO.\cite{harris,balakir}  
The high--temperature part of the data, from about 70 K to 300 K, 
is convex and as discussed
in Refs.[\cite{harris,balakir}] follows the $T$--dependence
of the square of the Hall angle in accord with Eq.~3.
At lower temperatures, below about 70 K,
the lines deviate upwards. 
The deviation leads to an inflection point which gives the curves their
S--shape.

We have noted previously\cite{balakir} that the inflection point moves to
higher temperatures when the strontium fraction is decreased,
until in strongly underdoped LSCO the curves lose their S--shape, as
the region of validity of Eq.~3 moves to higher temperatures, beyond
the region of the measurements.
In contrast, we see that the zinc impurities do not affect the shape
of the curves. They retain their S--shape, at least
in all superconducting specimens, showing that Eq.~3 remains valid 
for all films above 70 K.
We have suggested previously that the position of the inflection point
is related to the pseudogap opening. We will return to discuss this
point in the next section, but concentrate first on the
high--temperatures regime where the OMR is
proportional to $\tan^2{\Theta_H}$.

\begin{figure}[ht]
\vspace*{-1.4cm}
\epsfig{file=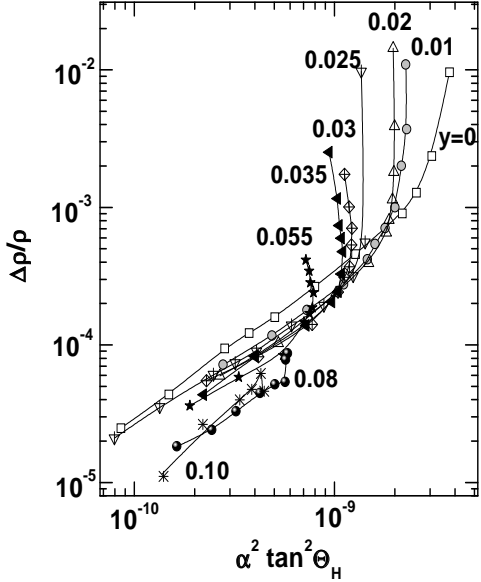, height=0.68\textwidth, width=0.42\textwidth}
\vspace*{-2.0cm}
\caption{The OMR as a function of ${\alpha^2}{\tan^2}{\Theta_H}$ for
a series of films with various values of $y$. The high--temperature data are 
on the left side of the figure. All lines are guides to the eye.}
\label{fig17}
\end{figure}

In order to investigate the relation between the Hall effect and the
OMR in more detail we show on Fig.~16a
the OMR data for the films with $y$ = 0 and $y$ = 0.025 as a function
of $\tan^2{\Theta_H}$.
Since the Hall angle is affected by strain and the OMR is not,
the data for specimens with the same $y$ and different $\rho_0$ fall
on separate curves. 
The high--$T$ region is to the left of the figure,
for small values of $\tan{\Theta_H}$, and small values of the OMR. The
dashed straight lines are fitted to the data in this regime. Within
experimental error the lines have the same slope, indicating the validity
of Eq.~3. As is true for $y$ = 0 (Fig.~2), the differences between the
data for different $\rho_0$ disappear when we plot the OMR against
$\alpha^2\tan^2{\Theta_H}$ (Fig.~16b).

Fig.~17 shows the data for all films with positive OMR as a function of
$\alpha^2 \tan^2{\Theta_H}$. We see that the high--temperature data
follow a set of parallel lines for all specimens, even
the nonsuperconducting ones. This shows that in this region the
OMR and $\tan^2{\Theta_H}$ have the same temperature dependence. The
straight lines shift downwards as $y$ increases, showing that the
coefficient of proportionality decreases as $y$ increases.

\begin{figure}[ht]
\vspace*{-1.0cm}
\epsfig{file=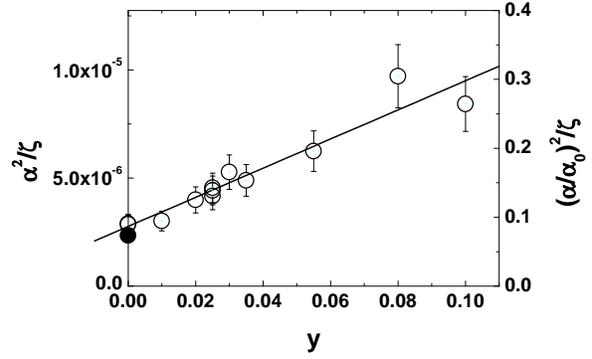, height=0.62\textwidth, width=0.43\textwidth}
\vspace*{-4.0cm}
\caption{The dependence of the reciprocal of the parameter
${\zeta}/{\alpha^2}$
on $y$. The scale on the right shows the values of the reciprocal of
${\zeta}({\alpha_0}/{\alpha})^2$. The full point is for the
La$_{1.83}$Sr$_{0.17}$CuO$_{4}$ single crystal of Harris {\em et al.},
Ref.[8].}
\label{fig18}
\end{figure}

We now rewrite Eq.~3 as

\begin{equation}
{\Delta}{\rho}/{\rho} = ({\zeta}/{\alpha^2})({\alpha^2}{\tan^2}{\Theta_H}).
\label{mrtansc}
\end{equation}

Since ${\alpha^2}{\tan^2}{\Theta_H}$ is strain--independent, and the
experiment shows that this is true also for the OMR, it follows that 
the coefficient ${\zeta}/{\alpha^2}$ 
is also not affected by strain and depends only on $y$.
Moreover it has a surprisigly simple dependence on $y$. In Fig.~18 we
plot its reciprocal as a function of $y$. This function is seen to
incease linearly with $y$. Since $\alpha$ is $y$--independent, this graph gives us 
also the $y$-dependence of $\zeta$. We can write

\begin{equation}
\frac{\zeta}{\alpha^2} = \frac{1}{K_0 + K_1 y},
\label{zet}
\end{equation}

\noindent
where  $K_0 = (2.75 \pm 0.16) \times 10^{-6}$ K$^{-4}$ and  $K_1 =
(6.54 \pm 0.55) \times 10^{-5}$ K$^{-4}$. 

The last column of Table 1 shows ${\zeta}/{\alpha^2}$ 
multiplied by ${\alpha_0}^2$, 
where $\alpha_0$ is the value of $\alpha$ for the film with $y = 0$ and with
the lowest residual resistivity, $\alpha_0 = 5.65 \times 10^{-3}$
K$^{-2}$, so that it is equal to $\zeta$ for $y$ = 0. It changes 
from 11.05 in the film with $y = 0$ to 3.78 in the film with $y = 0.10$.
We show the reciprocal of this quantity on the right--hand side of Fig.~18.
The value of $\zeta$ for the La$_{1.83}$Sr$_{0.17}$CuO$_{4}$ single crystal of 
Ref.[8] is 13.6, in good agreement with the film value.

The coefficient $\zeta$ in LSCO is substantially larger then in
other high-$T_c$ compounds. In YBCO $\zeta$ is equal to $1.5-1.7$,\cite{harris}
and in optimally doped and overdoped Tl$_2$Ba$_2$CuO$_{6+\delta}$ it is
equal to 3.6 and 2.0, respectively.\cite{tyler} Since the magnitude of the 
cotangent does not differ significantly between the different compounds, the ratio
seems to depend mainly on the magnitude of the OMR, which 
reflects the deviation of the Fermi surface from sphericity. The larger
$\zeta$ in LSCO would than be a natural consequence of the fact that the Fermi
surface is more flat in LSCO than in any of the other high--$T_c$
compounds. The addition of impurities reduces $\zeta$ 
to a value comparable to that of the 
thallium compounds.\cite{tyler} This is probably a consequence of the
reduction of the anisotropy caused by the addition of isotropic
impurity scattering. Since the magnetoresistance 
is an effect of higher order in magnetic field than the Hall effect, it should 
be more sensitive to the reduction of anisotropy than the Hall angle,
leading to a decrease
of $\zeta$ with $y$.

\subsection {Pseudogap}

We now return to the discussion of the inflection point, and its relation
to the pseudogap. 

\begin{figure}[ht]
\vspace*{-0.3cm}
\epsfig{file=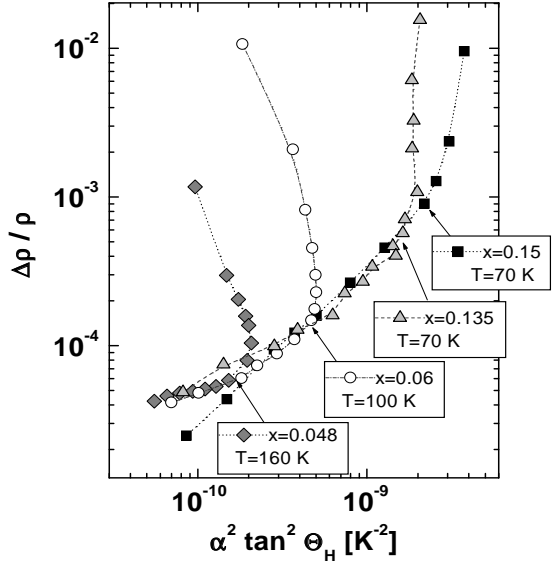, height=0.73\textwidth, width=0.53\textwidth}
\vspace*{-4.0cm}
\caption{The OMR as a function of ${\alpha^2}{\tan^2}{\Theta_H}$ for
La$_{2-x}$Sr$_{x}$CuO$_{4}$ films with various amounts of strontium, $x$,
as indicated in the figure. The dotted lines are guides to the eye. The
arrows identify the onsets of the saturation of of $\tan{\Theta_H}$, at
temperatures indicated next to each curve.}

\label{fig19}
\end{figure}

The inflection point marks a crossover from the high--$T$ regime,
where the OMR is proportional to $\tan^2{\Theta_H}$, to low--$T$ regime, where
this proportionality does not hold. The inflection
point is very easy to identify from Fig.~17, even in the case of
nonsuperconducting specimens. 
It is about 60 to 70 K, regardless of $y$. This is very different from
its shift in underdoped LSCO,\cite{balakir} which we show on Fig.~19.
Here the region where the data follow straight lines moves gradually 
to higher temperatures, and the slope of the line changes, showing that
Eq.~3 is not followed. This is consistent with our previous
finding that the $T$--dependence of the OMR looses it's S--shape in
underdoped LSCO because the inflection point moves to temperatures
outside the range of the measurements.\cite{balakir} 

\begin{figure}[ht]
\vspace*{-1.1cm}
\epsfig{file=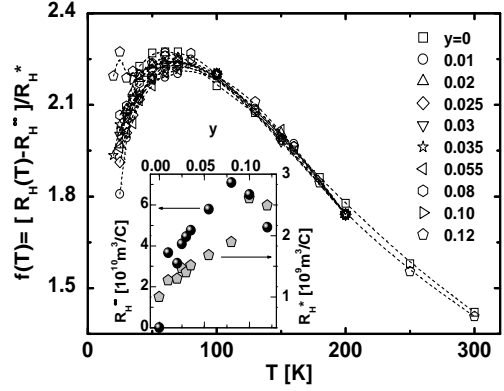, height=0.70\textwidth, width=0.45\textwidth}
\vspace*{-5.0cm}
\caption{The rescaled Hall coefficient, $R_H (T)-{R_H}^{\infty})/R_H^{*}$,
as a function of temperature. The inset shows the behavior of
the parameters
${R_H}^{\infty}$ and $R_H^{*}$ as a function of $y$.}
\label{fig20}
\end{figure}

When the loss of proportionality between the OMR and 
$\tan^2{\Theta_H}$ was first discovered, it was atributed to 
superconducting fluctuations, which would enhance the OMR 
in the vicinity of $T_c$.\cite{harris,kimura} A careful examination of
Figs.~17 and 19 shows, however, that the deviation of the two
quantities is related to the suppression
of $\tan{\Theta_H}$ rather than to an enhancement of the OMR. This is
most easily seen in the specimens 
with low $T_c$ and in nonsuperconducting specimens, in which the 
superconductivity does not get in the way.
The OMR continues to grow steadily at low temperatures, while $\tan{\Theta_H}$ 
first saturates, and then decreases. The temperature of the onset
of saturation does not follow the decrease of $T_c$, but remains constant
at about 60 to 70 K in the specimens with zinc, and increases in the
underdoped samples, as shown by arrows in Fig.~19. The only specimen in
which the enhancement of the OMR may play a role is the one with the optimal $T_c$,
presumably as a result of the close vicinity of the transition
to the superconducting state. 

A second effect, seen on Fig.~15 is a gradual decrease of the OMR,
below about 30 K in the film with $y = 0.08$, and below about 55 K in
the film with $y = 0.1$. This effect is $y$--dependent and quite 
distinct from the $y$--independent suppression of $\tan{\Theta_H}$.
As we mentioned before, it may be linked to 
localization effects which eventually lead to a negative OMR in the film with 
$y = 0.12$. We conclude that in the films with zinc there are two
distinct anomalies.
One, at higher $T$, is the suppression of $\tan{\Theta_H}$,
while the OMR continues to increase as $T$ is lowered. 
The second effect, at lower $T$, is a suppression of the OMR,
associated with localization effects. 

The suppression of $\tan{\Theta_H}$ at low $T$ has been observed before
in underdoped YBCO\cite{xu,abe} 
and in single--layer and bilayer bismuth compounds\cite{kons} 
suggesting that they may be related to the pseudogap opening. 
In Ref.[52] it was noted that 
the suppression of $\tan{\Theta_H}$, and the broad maximum in $R_H$ both occur at 
approximately the same temperature as the anomaly in the $^{63}$Cu NMR 
relaxation rate. In the Bi--2212 phase, on the other hand, 
the tangent anomaly occurs at lower $T$ than the anomalies 
in the NMR relaxation rate.\cite{kons}

\begin{figure}[ht]
\vspace*{-2.8cm}
\epsfig{file=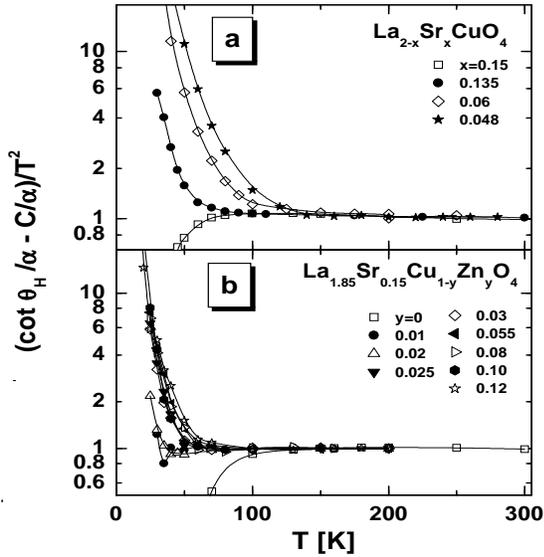, height=0.70\textwidth, width=0.55\textwidth}
\vspace*{-2.5cm}
\caption{The rescaled cotangent of the Hall angle,
$(\cot{\theta_H}/{\alpha}-C/{\alpha})/T^2$,
as a function of temperature, for several underdoped LSCO films (a), and
for a series of films with zinc (b).}
\label{fig21}
\end{figure}

\begin{figure}[ht]
\vspace*{-0.8cm}
\epsfig{file=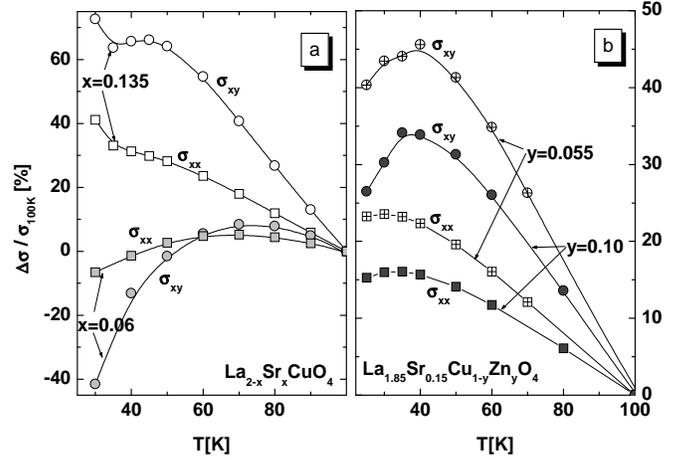, height=0.73\textwidth, width=0.5\textwidth}
\vspace*{-5.5cm}
\caption{The relative change of the longitudinal and Hall
conductivities, normalized at $T = 100$ K. The data are for two
underdoped LSCO films with $x = 0.135$ and $x = 0.06$ (a), and
for two films with $y = 0.055$, and $y = 0.06$ (b).
The Hall conductivity is measured at 8 tesla. All lines are guides to the eye.}
\label{fig22}
\end{figure}

The correlation between the maximum in $R_H$ and the suppression of 
$\tan{\Theta_H}$ in LSCO may be observed directly from the $T$--dependence 
of these quantities, which we now examine in more detail.
We have rescaled the Hall coefficient, $R_H$, by
the procedure outlined in Ref.[16], where it is shown 
that $R_H (T)$ 
for various values of $x$ collapses to a single curve if plotted 
as $(R_H (t)-{R_H}^{\infty})/R_H^{*}$, versus
$t=T/T^*$. Here ${R_H}^{\infty}$ is the asymptotic value
of $R_H$ at high $T$, while $R_H^{*}$ and $T^*$ rescale the Hall coefficient
and the temperature, respectively. $T^*$, which identifies the
temperature above which the Hall coefficient becomes $T$--independent,
was found to decrease from about 700 K for $x = 0.1$ to 100 K in
the overdoped regime, and was proposed later to be related to the
opening of the pseudogap.\cite{batlogg} We apply the same procedure to
the data of Fig.~10 for a series 
of films with zinc, and show the result on Fig.~20. The parameters of
the rescaled Hall coefficient are shown in the inset. We find that it
is not neccesary  to rescale the temperature, as could be anticipated 
from the fact that the maximum in $R_H (T)$ does not shift with $y$.
Comparison with Ref.[16] leads to $T^* \sim 600\,$K, 
independent of $y$, and close to the result of Ref.[16]. 

In order to examine
the deviation from Eqs. 2 and 5, we show on Fig.~21 the quantity
$(\cot{\Theta_H}/{\alpha}-C/{\alpha})/T^2$.
The top panel shows the data for 
several underdoped LSCO specimens, while the bottom contains the data
for the films with zinc. The data for the optimally doped LSCO
specimens deviate 
downwards in both figures, as a result of the close vicinity of $T_c$.
On the other hand, the data for both underdoped LSCO, and  LSZNCO,
deviate upwards, because of the suppression of $\tan{\Theta_H}$. 
The point of deviation clearly shifts to 
higher $T$ with the decrease of $x$, but remains 
constant, at about 50 to 70 K  when $y$ changes.

To determine the origin of the suppression of $\tan{\Theta_H}$ we have 
analyzed the $T$--dependence of the longitudinal and the Hall
conductivities. We find that in the LSCO films both conductivities decrease
at low temperatures, unlike the situation in 60 K YBCO,\cite{xu} 
where the suppression was found to be present only in the Hall conductivity.
This behavior is illustrated in Fig.~22, were we show the relative change of 
conductivities for two underdoped, and two films with zinc, normalized to $T = 100$ K. 
The onset of superconductivity is evident at the lowest temperature in
the film with $x = 0.135$. Apart from this all data display broad maxima in the 
conductivities, which shift to higher temperatures with decreasing $x$, and 
remain constant when $y$ changes, as for the anomalies in $\tan{\Theta_H}$.
However, in each specimen the Hall conductivity displays more pronounced 
anomalies, and at slightly higher temperatures than the longitudinal conductivity. 
This pattern indicates that the relatively stronger
changes of the Hall conductivity are the cause for the tangent suppression.
If we link this effect to the development of the normal--state gap,
our analysis reveals two important results. First, we conclude that the pseudogap 
opening affects the Hall conductivity more effectively than the longitudinal 
conductivity. Secondly, the pattern of the evolution with $x$ and $y$ indicates 
that the temperature at which pseudogap opens is not affected by the zinc doping,
while it shifts to higher temperatures for underdoped specimens. 

Finally, we would like to comment shortly on the decrease of the OMR
at low temperatures, which we propose to link to the localization effects.
These are not usual effects of weak localization observed in conventional 
disordered metals. We have made a careful study of the behavior of magnetoresistance 
of the zinc--doped and the underdoped LSCO at temperatures below these discussed 
in the present paper, at $T < 25\,$K , and the detail account will be published 
separately.\cite{artur} The results reveal that the weak localization effects are 
absent. Instead, we find strong evidence of the influence of spin--disorder scattering 
on the transport. This is not surprising. In LSCO the carriers move in the 
disordered magnetic background of the CuO$_2$ planes, and themselves
influence the type of this disorder, contributing to the $T$-dependence
of the resistivity. There is plenty of evidence that the 
localization of carriers is non-uniform in LSCO,\cite{julien1,julien2,ino1,ino2} 
possibly in form of stripes.\cite{ke} It is important to point out that
these localization effects appear to be distinctly different from the effects of 
the normal--state pseudogap opening which are observed in the high--temperature
range.

\section {Discussion}

The main subject of this study, which is the influence of the impurity doping 
and strain on the normal--state transport properties, may be summarized as 
follows.

{\em (1) Strain}

If we exclude the effect of grain--boundary scattering, the strain 
affects $\rho_0$, $A$, $\alpha$ and $C$ in approximately the same way.
A decrease of the $c$--axis lattice parameter, accompanied by an
increase of the $a$--axis parameter 
(i.e. tensile in--plane strain), results in an increase of all
transport parameters, 
with a fractional changes of about 65\% per 0.01 {\AA }. 
The OMR, on the other hand, is not affected by strain. 

{\em (2) Impurities}

The addition of zinc does not change the carrier concentration. It adds
an impurity scattering term to the residual resistivity and to the
elastic Hall scattering term. It has almost no effect on $A$
and does not affect $\alpha$. These results are consistent
with previous observations. However, the OMR is more strongly
suppressed by impurities
than ${\tan^2}{\Theta_H}$, so that $\zeta$ decreases
with an increase of impurities.

Before we compare these results to more sophisticated theoretical models, 
we compare them first to a simple Drude model, where the
resistivity is given by $\rho = {m^*}/n{e^2}\tau$, 
$n$ is the carrier concentration, $m^*$ the effective mass,
and $\tau^{-1}$ the relaxation rate, with both an elastic 
and an inelastic part. The Hall effect and the magnetoresistance are described 
by the relations: ${\cot}{\Theta_H} \sim ({\omega_c}{\tau})^{-1}$,
and ${\Delta}{\rho}/{\rho} \propto ({\omega_c}{\tau})^2$, where $\omega_c = eB/m^*$
is the cyclotron frequency.
In our experiment the tensile strain, which increases the in--plane distances,
causes simultaneous proportional increases in all transport coefficients, i.e.
it changes both the elastic and inelastic terms in $\rho$ and $\cot{\Theta_H}$
in the same way. In the Drude model this can only be explained by an
increase of the effective mass, which would qualitatively be a reasonable 
outcome of the increase of the in--plane lattice distances. However, we would
expect the OMR to change as well, since it is 
proportional to ${\omega_c}^2$, and this is not observed. In addition, 
$\zeta$ depends on the impurity content, while it should be
constant in this model.

We now consider more complicated approaches, designed to reproduce
the different $T$-dependences of $\rho$ and $\cot{\Theta_H}$.
We start with the non--FL models. The original model by 
Anderson\cite{anders}
invokes a picture of holons and spinons which separately control the
longitudinal and transverse relaxation rates, respectively. A subsequent 
model, also with two distinct relaxation rates, assumes that the electronic 
scattering in cuprates is sensitive
to the charge--conjugation symmetry of quasiparticles.\cite{col} The
main feature of both models is that two relaxation rates, which are a result
of spin--charge separation or different parity of the quasiparticle states,
exist at each point of the Fermi surface. 
Since in these models the Hall effect and magnetoresistance are
goverened by the same relaxation  
rate, their ratio should behave as in the 
Drude model, i.e. the coefficient $\zeta$ should be independent of temperature,
impurities, or strain. Our results are inconsistent with these expectations.

The FL models assume the existence of quasiparticles with strongly anisotropic 
scattering rates along the Fermi surface. Various anisotropies have
been proposed, 
including ``hot spots'' and ``cold spots'', small regions of Fermi 
surface in which the scattering is either much stronger, or much weaker, 
respectively, than in the remaining parts, and has a distinctly
different $T$--dependence.\cite{carr,kend,stoj,hlub,ioffe,zhel1,zhel2} 
Possible microscopic origins of different scattering rates include
antiferromagnetic spin fluctuations, charge fluctuations, and pairing fluctuations 
which may couple preferentially to carriers with certain 
momenta.\cite{stoj,ioffe,caprara} 
These concepts are based on ARPES results which indicate that single 
particle scattering is much stronger along the (0,0)--($\pi$,0) and 
(0,0)--(0,$\pi$) directions than along the zone diagonals.\cite{shen} 
Since $\rho$ and ${\cot}{\Theta_H}$
involve scattering in different regions of the Fermi surface,
it is then possible to reproduce the different $T$--dependences of $\rho$ and 
$\cot{\Theta_H}$.

The magnetoresistance has 
been evaluated in only two of these models.\cite{ioffe,zhel2} One of them
is the cold--spots model of Ioffe and Millis\cite{ioffe}, which assumes
that the in--plane 
resistivity is controlled by carriers with momenta along the zone diagonals,
which have an FL scattering rate proportional to $T^2$ (cold spots).
In the regions away 
from the diagonal, the scattering rate makes a large $T$--independent but
momentum--dependent contribution. The linear $T$--dependence of the resistivity 
results from the linear $T$--dependence of the width of the cold region.
This model has been used sucsessfully to explain several 
experiments.\cite{marel,hardy} However, the magnitude of the OMR 
is much larger than experimentally observed and the coefficient $\zeta$ is found 
to be $T$--dependent. These features disagree with both the previous experiments,
and with the present study. The model
predicts a violation of Matthiesson's rule and
this is also contrary to the experimental data. In addition, the ratio
$\zeta$ should increase with $y$, while our experiment indicates a decrease. 

A second calculation of the OMR is based on the phenomenological additive 
two-$\tau$ model, which assumes two distinct relaxation rates $\tau_1$ and 
$\tau_2$.\cite{zhel1,zhel2} The Fermi surface is assumed to have large flat 
regions around $M$ points with a short relaxation time $\tau_1 \sim T^{-1}$ and large 
Fermi velocity (hot spots). The sharp corners around the nodal points have 
a long relaxation time $\tau_2 \sim T^{-2}$ and small Fermi velocity (cold spots). 
The carriers from the hot spots dominate the in--plane resistivity, while cold 
regions dominate the Hall conductivity, leading to $T$--dependences which 
agree with experiment. This model leads to an almost $T$--independent ratio $\zeta$, 
close to the experimental results. The impurity effects on $\zeta$ have not been 
evaluated.
ARPES experiments on bismuth compounds do not confirm the assumed character of
the Fermi surface.\cite{camp} The flat portion around the $M$ points is 
observed to be smaller, and the velocities in the hot region are smaller than 
in the cold region. Interestingly, the large flat portions around $M$
may better approximate the real Fermi surface observed in LSCO. 

Apart from these comparisons the effect of strain 
creates the most stringent test of the FL models.
The models have to reflect the fact that strain affects all transport coefficients 
similarly, but does not affect the OMR. The effect of strain on the Fermi
surface can change the relative size of the cold and hot areas.
In the two--$\tau$ model the ratio of the contributions from the cold and
hot areas appears to be approximately the same for the Hall conductivity 
and for the magnetoresistance, so that the different effects of strain
on these quantities may be difficult to reproduce. However, a detailed comparison
is needed to evaluate this effect.

A recent ARPES study provides more detail on the properties
of the Fermi surface.\cite{valla} It finds
that in Bi$_2$Sr$_2$CaCu$_2$O$_{8+\delta}$ (Bi2212)
the single--particle scattering rate contains a large $T$--independent
part which disappears only in the vicinity of the nodal directions,
plus a part linearly dependent on temperature and energy which extends
over most of the Fermi surface and becomes almost $T$--independent in the
vicinity of the (${\pi},0$) and ($0,{\pi}$) directions. This result
motivated the development of two new theoretical approaches.\cite{va,PSK}

The model of Varma and Abrahams combines the predictions of the marginal 
FL hypothesis for the inelastic scattering linear in $T$ with an elastic, strongly 
anisotropic term, which results from small--angle 
forward scattering by impurities situated away from the CuO$_2$ planes.\cite{va} 
The forward scattering produces new
contributions to $\tan{\Theta_H}$ so that
instead of Eq.~2 the resistivity should be proportional to
$\sqrt{\cot{\Theta_H}}$. The data for single
crystals of YBCO with zinc seem to support this form.\cite{va}
A preliminary analysis of our data shows that this proportionality is followed
for a film with $y = 0$ and small strain. However, deviations appear
for films with strain.
The proposal of Ref.[32] has also been 
discussed by Hlubina\cite{hlub2} who argues that the new contributions to 
$\cot{\Theta_H}$ from forward scattering will introduce terms of the same 
magnitude in the resistivity as well, so that the differences between
the $T$--dependences 
of $\rho$ and $\cot{\Theta_H}$ may not be observable. 

Another model introduced recently is the two--patch model, designed to analyze
normal state transport properties of cuprates using the Boltzmann equation.\cite{PSK}
The Brillouin zone and the Fermi surface are divided into regions where 
the scattering between the electrons is strong and the Fermi velocity is 
low (hot patches), and regions where the scattering is weak and the Fermi velocity 
is large (cold patches). For Bi--based cuprates the hot patches are centered 
around the saddle ($M$) points of the Brillouin zone, while the cold patches are centered 
around the nodal points, along the $\Gamma Y(X)$ direction of the Brillouin zone.
Three distinct temperature dependences for the scattering amplitude are assumed,
$T^2$ in the cold region, $T$ for the inter--patch (hot--cold) scattering,
and a $T$--independent value in the hot region. The resulting scattering amplitude 
$1/\tau_{k}$ obtained from the scattering matrix is strongly momentum dependent 
and the low--temperature behavior, in contrast with other similar 
approaches,\cite{hlub,ioffe,zhel1,zhel2} is always non--FL, with a linear 
$T$--dependence in the cold patches and a constant in the hot patches, as observed 
by the ARPES spectra.\cite{valla} This model gives a reasonable description 
of the transport properties of the $Bi$--based cuprates. When comparing it
to our experiment we note first that the model predicts an increase of the residual 
resistivity and of the constant term in the cotangent when
the size of hot regions increases. Since hot regions may be expected to
be larger in the flatter
Fermi surface of LSCO, this would explain why the residual resistivities 
and the constant term in the cotangent are larger in LSCO than in other cuprates. 
In addition, the $T$--dependence of the longitudinal and the Hall conductivities 
in this model are both controlled primarily by the density of states
and the Fermi velocities in the cold regions. The similar effect 
of strain on these quantities could then be explained by the effect of strain
on the cold--region properties. It is not clear if this model can reproduce
the insensitivity of the OMR to strain. To see this 
more detailed calculations using the LSCO Fermi surface would be necessary.

To summarize this part of the discussion, we conclude that our results disagree
with most of the earlier theoretical models.
New theoretical models,\cite{va,PSK} which include the Fermi surface
properties in a more realistic way may turn out to be more
compatible with experiment. More detailed comparisons are needed to evaluate these 
new proposals, including the effects of strain, as described in this study.
 
Next we comment on the 
markedly different effect of a change of $x$ (in underdoped LSCO),
and $y$ (in zinc--doped LSCO) on the pseudogap, as inferred from the
suppression of $\tan{\Theta_H}$.

The gradual opening of a pseudogap was originally suggested to explain
anomalies in the behavior of various normal--state properties.\cite{timusk}
These anomalies take the form of a crossover temperature between two
different $T$--dependences. An example is the Hall effect anomaly, where
below the crossover temperature, $T^*$, the  Hall coefficient is
$T$--dependent.\cite{hwang}  The crossover temperatures 
for different properties and materials differ a great deal. However, most of them
increase with underdoping. 
For example, in LSCO $T^*$ increases from 100 K to 700 K 
when $x$ is decreased from 0.3 to 0.05.\cite{hwang}

ARPES studies provided a number of important insights into pseudogap phenomena. 
The opening of the gap in the normal--state excitation spectrum above $T_c$ 
has been detected in underdoped Bi2212 and Bi2201. 
Its momentum dependence is consistent with $d$--wave symmetry, and as the
temperature decreases the normal--state gap evolves smoothly into a superconducting 
gap.\cite{ding,loeser,harris3,white,norman,ron} The temperatures of the
pseudogap opening from tunneling, optical conductvity, and  Raman scattering 
experiments, are in quite good agreement with the ARPES
data on Bi2212.\cite{timusk}

The situation is more confusing for LSCO. The photoemission studies 
are more difficult because of quick surface degradation at high temperatures. 
The ARPES studies were limited to low temperatures,\cite{ino1,ino2} while 
angle--integrated photoemission spectra (AIPES) were investigated as a 
function of $T$.\cite{tsato} An AIPES study of optimally--doped LSCO reveals a 
suppression of the density of states near the Fermi energy as $T$ is lowered,
extending to about 30--35 meV.\cite{tsato}  This energy corresponds
to about 350 to 400 K, which is somewhat lower than the Hall--effect
crossover temperature $T^*$ in optimally doped LSCO.\cite{hwang} However, 
no evidence was found of the connection between this pseudogap and the 
superconducting gap. 
The ARPES study focuses on the dependence on $x$ of the spectra
measured around the saddle point ($\pi, 0$) 
and concludes that the energy gap, $\Delta$, increases smoothly with
decreasing $x$. In an optimally--doped crystal $\Delta$ is a
superconducting gap of about 8 meV, while in an underdoped specimen
with $x = 0.05$ it is a normal--state gap of about 25 meV.\cite{ino1,ino2} 
In addition, the spectral weight around the nodal
points becomes severely depleted below $x = 0.12$, while the band
around the saddle points is very flat. 

In order to correlate the suppression of $\tan{\Theta_H}$, observed in our 
measurements, with the other anomalies,
we compare their $x$--dependences on Fig.~23.  The full circles show
the temperature $T_{tan}$ at  
which $\tan{\Theta_H}$ is suppressed for the underdoped LSCO films from Fig.~19. 
As we discussed earlier, in optimally--doped films the vicinity
of the superconducting state prevents the suppression of $\tan{\Theta_H}$ from 
being seen. Instead we plot the temperature  
of the inflection point for $x = 0.15$ and $x = 0.225$. 
We also plot $T_c$ for all films. Comparing the two dependences we see that the
line describing the suppression of $\tan{\Theta_H}$ is considerably
higher than the $T_c$--line in the underdoped films, but
approaches $T_c$ in the optimally doped films.

\begin{figure}[ht]
\vspace*{-1.8cm}
\epsfig{file=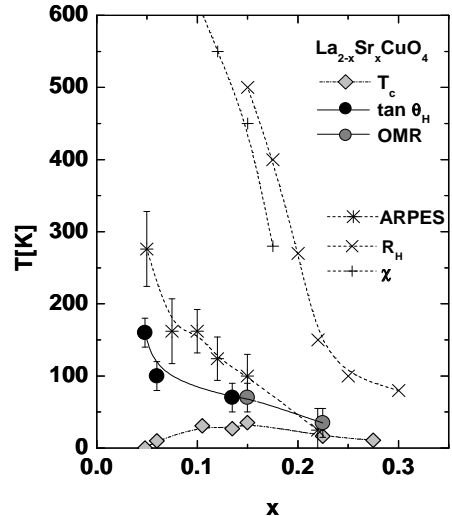, height=0.73\textwidth, width=0.52\textwidth}
\vspace*{-4.5cm}
\caption{The $x$-dependence of several quantities for La$_{2-x}$Sr$_{x}$CuO$_{4}$. 
Circles: temperature of the tangent suppression (black), inflection
points from OMR data (grey); diamonds: $T_c$; stars: $\Delta$
from ARPES, Ref.[56]; crosses: anomalies from Hall effect and susceptibility,
Ref.[16]. All lines are guides to the eye.}
\label{fig23}
\end{figure}

Next we include the crossover temperatures of the Hall effect ($T^*$), and
the susceptibility from  Ref.[\cite{hwang}], and the gap from the
ARPES experiments.\cite{ino2} Our results for $T^*$ give a value of about 600 K,
which is in reasonable agreement with the $T^*$--line in the figure. 
It is clear that $T_{tan}$ is considerably smaller than $T^*$. Instead, it seems
to be quite closely related to the normal--state gap value inferred from the ARPES
experiments. Therefore we conclude that the Hall conductivity in the underdoped
films is strongly affected by the opening of the normal--state gap around the 
saddle point ($\pi$,0).
 
It is important to point out here that this conclusion does not nessecerily 
mean that the nodal excitations are unimportant for the normal--state Hall effect. 
As we discussed, in the underdoped specimens Eq.~3, characteristic for the
normal state of the optimally doped films, is not fulfilled. This indicates
that in the underdoped samples the density of states which contributes
to the transport properties is already
severely affected at high temperatures, and this is
reflected in the shift of $T^*$ with underdoping. While
this effect may be a precursor of the normal--state gap opening, the ARPES
results suggest that there is another possibility, related to the decrease
of the spectral weight around nodal points in the underdoped specimens.\cite{ino1,ino2}
It is entirely possible that in the optimally--doped range the nodal density of states
contributes decisively to the transport propertis, and after it is
eliminated by the 
underdoping the only contribution which is left is from the saddle point. 
Alternatively, the effect may be related to the shrinking of the size of 
Fermi surface regions without a gap around the nodal points. This would be more
in accord with the fact that a similar coincidence between $T_{tan}$ 
and the normal--state gap opening deduced from ARPES occurs for bismuth 
compounds,\cite{kons,ding} in which there is no substantial decrease of the 
spectral weight around nodal points in the underdoped compositions. This would
indicate that the origin of the effect of normal--state gap opening on the
suppression of $\tan{\Theta_H}$ is the same for different cuprate compounds. 

Finally we discuss the effect of zinc on the pseudogap. Our experiment
shows that both crossover temperatures, $T^*$, and $T_{tan}$, remain constant
when $y$ changes, indicating that the opening of the normal--state 
gap is unaffected by the zinc. On the other hand, the magnitudes of both the OMR
and $\tan{\Theta_H}$ decrease with $y$, with the OMR affected more strongly as
shown by Eq.(8). These two seemingly contrasting results, i.e. the constant
temperature of the gap opening, and the influence of impurities on the
transport properties
may be reconciled if one assumes that the effect of impurities on the
pseudogap is confined to the immediate area surrounding the impurity, while
away from them the pseudogap remains intact. In fact, there have been many
suggestions that this is indeed the case. $^{63}$Cu NMR experiments
on Zn-doped YBCO find an enhancement of the antiferromagnatic correlations around
Zn impurities, while the crossover temperature in the relaxation rate
remains constant.\cite{julien1} Similar conclusions have been inferred from
neutron scattering experiments\cite{neutrons}, from
$^{89}$Y NMR measurements,\cite{mahajan} and ESR measured on gadolinuim sites
in Gd-, and Zn-doped YBCO.\cite{janos} A local effect of zinc impurities
on the pseudogap has also been suggested by studies of thermopower\cite{tallon} 
and specific heat in YBCO.\cite{loram} 

These local effects resemble the ``swiss cheese'' model,\cite{nach}
in which charge carriers around each Zn impurity are excluded from
superconductivity. However, the real effect of impurities on the normal state properties
is far more complex. While the temperature of the pseudogap opening is unaffected 
in the main volume of the sample, the other properties are strongly affected. 
These effects are very different from those in
underdoped LSCO, as can be seen from the fact that the main features of the 
normal state, given by Eq.~1 to 3, survive in the films with zinc.

\section{Conclusions}

The analysis of the structure and microstructure of the
La$_{1.85}$Sr$_{0.15}$Cu$_{1-y}$Zn$_y$O$_{4}$ films shows that they
grow with variable amounts
of built--in strain resulting from the partial relief of the lattice
mismatch by dislocations. Both compressive and tensile in--plane strain 
with respect to the bulk lattice parameters are observed. They are accompanied 
by expansion or compression of the $c$-axis lattice parameter, respectively.
Grain--boundary scattering enhances the
residual resistivity in the regime
of large tensile in--plane strain, but does not affect the other
transport parameters.

Strain affects the superconducting and the normal--state transport
parameters. $T_c$  
decreases with the decrease of the $c$--axis lattice parameter 
at a rate of about 680 K/{\AA}. The decrease of $T_c$ is accompanied by a
linear increase of $\rho_0$, $A$, $\alpha$ and $C$ at a rate of 8.3\% per Kelvin.
The addition of zinc adds an impurity scattering term to $\rho_0$, and
to $C$, the
constant term in $\cot{\Theta_H}$, while the slope
of the $T$--dependence of the resistivity and the slope of the $T^2$--dependence
of the cotangent remain unchanged. The effects of impurities and strain 
on $\rho_0$ are additive, while they are multiplicative in the case of $C$.
The OMR is independent of strain. Over a limited $T$--range, above the
inflection point,  
the OMR is proportional to ${\tan^2}{\Theta_H}$.
The coefficient of proportionality, $\zeta$, dependens on strain and on 
the impurities, showing that the relaxation rate which
governs the Hall effect is not the same as that of the magnetoresistance.
A comparison of these results with the available theoretical models of the
normal state indicates that none of them can fully describe the 
experiments. New models which take the properties
of the Fermi surface into account more realistically may be compatible,
but a more detailed
evaluation, including the strain effects which we observe will be necessary.

In addition, we observe a suppression of $\tan{\Theta_H}$ for underdoped
and zinc--doped films, and show that it can be associated with the
opening of a gap in the normal--state excitation spectrum. 
The temperature of the pseudogap opening does not change with the
addition of zinc impurities
but it increases when the Sr--La ratio is decreased.
At temperatures lower than the temperature of the pseudogap opening, 
the OMR decreases as a result of localization effects. 

\acknowledgements
We would like to thank M. Gershenson and S-W. Cheong for their cooperation 
and sharing of laboratory facilities. We also thank T. E. Madey for providing the
facilities for the AFM study and for helping with its interpretation, and C. L. Chien 
for the 4--cycle X--ray diffractometer measurements. 
We appreciate the discussions with Piers Coleman, Gabriel Kotliar, and Andrew Millis. 
We also take the opportunity to thank Richard Newrock and 
the Physics Department of the University of Cincinnati for help with the 
construction of the target chamber. 
This work was supported by the Polish Committee
for Scientific Research, KBN, under grants 2 P03B 09414 and 7 T08A 00520,
by the Naval Research Laboratory, by the Rutgers Research Council,
and by the Eppley Foundation for Research. A. Perali acknowledges partial
support from Fondazione "Angelo della Riccia".

\end{document}